\documentclass[prd,a4paper,nofootinbib, preprintnumbers, amsmath,multicol,12pt]{revtex4}

\usepackage{color}
\usepackage{latexsym}
\usepackage{epsfig}
\usepackage{amsmath}
\usepackage{amssymb}
\usepackage{graphicx}
\usepackage{bm}

\makeatletter

\newcommand\ee{\end{equation}}
\newcommand\be{\begin{equation}}
\newcommand\eea{\end{eqnarray}}
\newcommand\bea{\begin{eqnarray}}

\def\beq{\begin{equation}}
\def\eeq{\end{equation}}

\begin{document}
\setcounter{page}{0}
\thispagestyle{empty}

\hfill UAB-FT-741\hspace{3.2cm} CP3-Origins-2013-27 \hspace{3.2cm} DIAS-2013-27



~\vspace{1cm}

\begin{center}

{\Large\bf \color{red}  
On the Veltman Condition, the Hierarchy Problem and  High-Scale Supersymmetry}
\vspace{1cm}

{\large  
Isabella Masina$^{a,b}$, Mariano Quiros$^{c}$  }
\\[.5cm]
{\normalsize \small { \sl
$^{a}$ Dip. di Fisica e Scienze della Terra, Universit\`a di Ferrara and INFN Sez. di Ferrara, \\ \vspace{-.2cm}Via Saragat 1, I-44100 Ferrara, Italy}}\\
{\normalsize \small { \sl 
$^{b}$ CP$^\mathbf 3$-Origins and DIAS, Southern Denmark University, \\ \vspace{-.2cm}Campusvej 55, DK-5230 Odense M, Denmark}}\\
{\normalsize \small { \sl
$^{c}$ Instituci\'o Catalana de Recerca i Estudis  
Avan\c{c}ats (ICREA) and\\ \vspace{-.2cm}Institut de F\'isica d'Altes Energies, Universitat Aut{\`o}noma de Barcelona\\
\vspace{-.2cm} 08193 Bellaterra, Barcelona, Spain
}}\\
\end{center}
\vspace{.8cm}

\setcounter{page}{1}

\begin{abstract}\baselineskip=15pt
\begin{center} {\bf Abstract} \end{center}
In this paper we have considered the possibility that the Standard Model, and its minimal extension with the addition of singlets, merges with a high-scale supersymmetric theory at a scale satisfying the Veltman condition and therefore with no sensitivity to the cutoff. The matching of the Standard Model is achieved at Planckian scales. In its complex singlet extension the matching scale depends on the strength of the coupling between the singlet and Higgs fields. For order one values of the coupling, still in the perturbative region, the matching scale can be located in the TeV ballpark.  Even in the absence of quadratic divergences there remains a finite adjustment of the parameters in the high-energy theory which should guarantee that the Higgs and the singlets in the low-energy theory are kept light. This fine-tuning (unrelated to quadratic divergences) is the entire responsibility of the ultraviolet theory and remains as the missing ingredient to provide a full solution to the hierarchy problem. 
\end{abstract}

\maketitle 


\baselineskip=16pt

\vspace{.8cm}


\newpage

\section{Introduction}
\label{introduction}
In view of the recent discovery of a particle consistent with the Standard Model (SM) Higgs boson 
with a mass $m_H\sim$ 126 GeV, announced by the ATLAS~\cite{:2012gk}  and CMS~\cite{:2012gu} collaborations at CERN, the issue of quadratic divergences in the Standard Model Higgs self-energy gains interest. Indeed quadratic divergences are indicative of the fact that the natural order of magnitude of the Higgs mass is $\mathcal O(f_L\Lambda)$ where $f_L$ is a loop factor and $\Lambda$ represents the scale of new physics beyond the Standard Model (BSM).

Quadratic divergences in the Standard Model were studied by Veltman~\cite{Veltman:1980mj} in the context of dimensional regularization getting the one-loop condition~\footnote{Notice that the result corresponding to Eq.~(\ref{ecuacionVeltman}) was implicit in the early work of Ref.~\cite{DeckerPestieau}, through a tadpole diagram for the physical Higgs contributing to the quantity $\delta m_i^2/m_i^2$ $(i=q,\ell,W,Z,H)$ computed in the broken phase.} 
\beq
m_H^2+2 m_W^2+m_Z^2-4m_t^2=0
\label{ecuacionVeltman}
\eeq
which is satisfied for a value of the Higgs mass $m_H\sim 314$ GeV in flagrant conflict with experimental data.

Meanwhile BSM theories aiming to solve the problem of quadratic divergences have been postulated. There are essentially two class of such theories:
\begin{itemize}
\item
Theories where the Higgs is composite in the infrared by some strong dynamics. Therefore at high energies the Higgs dissolves into its constituents, only fermionic matter is present and there are no quadratic divergences. This solution is unrelated to the Veltman condition and we will not be concerned about it.
\item
Theories with extra fields and an extra symmetry such that the new fields with couplings dictated by the symmetry cancel the quadratic divergences of the Higgs self-energy. The prototype of such theories is supersymmetry and in particular the minimal supersymmetric extension of the Standard Model (MSSM). Unlike the previous solution the theory remains perturbative up to high scales. In a supersymmetric theory the absence of quadratic divergences is automatically satisfied~\footnote{Notice that Eq.~(\ref{ecuacionVeltman}) translates the absence of quadratic divergences only in the Standard Model. This equation is modified in extensions of the Standard Model as we will see later on in this paper. For examples in supersymmetric extensions cancellation of quadratic divergences is automatic as every term in Eq.~(\ref{ecuacionVeltman}) has a counterpart from its respective supersymmetric partner equal in absolute value and with different sign.}. In this paper we will consider this kind of solutions.
\end{itemize}

The search of supersymmetric particles is being the subject of an intense experimental activity at LHC~\cite{SUSYexp} although for the moment only negative results have been collected and only lower bounds on the mass of supersymmetric particles can be set. Most likely supersymmetry, if it exists at all, is only realized at high-scale.

In this article we will link both facts: the non appearance of supersymmetric particles at the LHC energies and the fact that the Veltman condition is not satisfied for the measured value of the Higgs mass. We will consider that the Veltman condition, although it is certainly not satisfied at the electroweak scales, can take place at some high-energy scale $\mu_V$ at which a supersymmetric theory takes over. Speculations on the possibility that the Veltman condition is not satisfied at low energy but at high scales have already been considered in earlier studies~\cite{Jack:1989tv,Alsarhi:1991ji,Chaichian:1995ef} when the Higgs mass was not known and thus no firm conclusions could be drawn. More recently  some authors have reconsidered the Veltman condition and pointed out~\cite{Holthausen:2011aa,Hamada:2012bp,Jegerlehner:2013cta,Jegerlehner:2013nna} that it is indeed satisfied close to the Planck scale.

The theory below $\mu_V$ is an effective Standard Model (or some minimal extension thereof) 
where the Veltman condition implies that there are no quadratic divergences. Therefore considering the SM as an effective theory valid for scales below $\mu_V$ the Higgs mass is not sensitive to the cutoff scale. Beyond $\mu_V$ the theory is assumed to be supersymmetric and thus the sensitivity to scales larger than $\mu_V$ is canceled by supersymmetry. Still the UV theory requires some fine-tuned relation among its parameters to match both theories at the scale $\mu_V$. This tuning is the remnant of the old SM hierarchy problem. Either it can have some environmental origin (e.g.~due to the huge number of vacua of the fundamental theory)  or perhaps it is guaranteed by some extra symmetry, or even it is a fine-tuning we have to live with (as it is the case of the cosmological constant problem). In any case the fine-tuning should be provided by the UV theory.

The paper is organized as follows. In section~\ref{QD} the problem of quadratic divergences in the Standard Model is reviewed along with the relation between the Veltman condition and the $Str\mathcal M^2$. In section~\ref{VC} the one-loop Veltman condition is numerically analyzed in the Standard Model using the NNLO running (three-loop for the beta functions and two-loop for the couplings matching) renormalization group equation. We show that for realistic values of the top quark and Higgs masses the Veltman condition is satisfied around Planckian scales. In section~\ref{MSSM} we study the merging of the Standard Model with the MSSM at the scale at which the Veltman condition is satisfied and with a predicted value of the parameter $\tan\beta$. Within the experimental errors in the top quark and Higgs masses and in $\alpha_3(m_Z)$ the value of $\tan\beta$ in the high-scale MSSM lies in the interval $1\lesssim\tan\beta\lesssim 2$. The merging of the Standard Model and the MSSM at the Veltman scale requires a fine-tuning which is the remnant of the hierarchy problem in the absence of quadratic divergences. In section~\ref{singlets} we study the Veltman condition in the presence of a complex singlet coupled to the Standard Model Higgs with coupling $\lambda_{SH}(\mu)$. Actually the scale at which the Veltman condition is fulfilled depends on the value of $\lambda_{SH}(m_Z)$. In fact for $\lambda_{SH}(m_Z)\simeq \mathcal O(1)$ the Veltman scale is $\mathcal O$(TeV). Moreover in order to fulfill the Veltman condition along the singlet field some massive vector-like fermions, coupled to the singlet, need to be introduced. These fermions can be Standard Model singlets in order to not perturb the running of the Veltman condition along the Higgs field. In section~\ref{NMSSM} the merging of the previous theory with the MSSM with the addition of singlet fields is done at the Veltman scale. We show that the merging with the supersymmetric theory cannot be done for any value of $\lambda_{SH}(m_Z)$. In fact it can only be done either for very small values of $\lambda_{SH}(m_Z)$ (essentially zero, i.e.~the Standard Model case) or for $\lambda_{SH}(m_Z)\gtrsim 0.3$. For given values of the parameters in the non-supersymmetric theory $\tan\beta$ in the high-scale supersymmetric theory is a prediction. The addition of the singlet requires an additional fine-tuning in the supersymmetric theory, on top of the MSSM's one, to insure the lightness of the Standard Model Higgs and a second one to satisfy the Veltman condition along the direction of the singlet. Finally section~\ref{sec-concl} contains our conclusions.

\section{Quadratic divergences in the Standard Model}
\label{QD}
According to~\cite{Veltman:1980mj} within the framework of dimensional regularization
a suitable criterion to address the issue of quadratic divergences is the occurrence of poles in the complex dimensional plane for $D$ less than four. In particular at the $n$-loop level, a quadratic divergence corresponds to a pole at $D=4-2/n$. Naive quadratic divergences at the one-loop level thus correspond to poles for $D=2$. 

Inquiring after the existence of poles for $D=2$ in the SM, it was realized  
by Veltman~\cite{Veltman:1980mj} that such poles exist in vector boson and Higgs self energy diagrams~\footnote{And also in tadpole diagrams and in connection with the cosmological constant.}. In particular for the Higgs mass they correspond to the shift, 
$m_H^2 \rightarrow m_H^2 +\delta m_H^2 $, and  the divergence has the 
form~\footnote{A similar structure holds for the vector bosons.}
\beq
\delta m_H^2=\frac{\Lambda^2}{16 \pi^2} C_{V}\, ,\, C_V= \sum_{n\geq 1} C_{Vn} ,
\eeq
where the contribution $C_{Vn}$  is associated to $n$ loops.
In particular at one-loop the Standard Model result is~\cite{Veltman:1980mj}
\beq
C_{V1} = \frac{3}{v^2} (m_H^2 +m_Z^2 +2 m_W^2 -4 m_t^2  ) \,.
\label{VC1}
\eeq

The condition for the absence of the quadratic divergences at one loop stemming from the cancellation between the fermion and boson masses, $C_{V1}=0$, is known as the Veltman condition at 1-loop (VC1)~\cite{Veltman:1980mj} and was dubbed "semi-natural" by Veltman himself.

A very simple way to understand the Veltman condition in the Standard Model and generalizations thereof is starting from the one-loop effective potential in the presence of the (constant) background Higgs field configuration $\phi$ 
\beq
V^{(1)}(\phi) = \frac{1}{64 \pi^2} \int d^4k \,\,Str \left[ \log(k^2+\mathcal M^2(\phi)) \right] 
\eeq
where
\beq
Str \left[ \log(k^2+\mathcal M^2(\phi)) \right] =  \sum_{J=0,\frac{1}{2},1} (-1)^{2J} (2J+1) \, Tr\left[   \log(k^2+M_J^2(\phi)) \right]
\eeq
and where $M^2_J(\phi)$ is the matrix of the second derivatives of the Lagrangian at zero momentum $k$ for spin $J$ fields~\footnote{Notice that for spin-1/2 fields one should replace $M_{1/2}^2(\phi)\to M_{1/2}^\dagger(\phi)M_{1/2}(\phi)$}.
The mass matrix is thus obtained from $M^2_J(\phi)$ by inserting the vacuum expectation value $\phi=v$, where $v$ is the location of the minimum of the effective potential.

The UV divergences of the one loop effective potential can be displayed by expanding the integrand in powers of large $k$.
Writing
\beq
\log(k^2+M_J^2) = \log k^2 + \frac{M^2_J}{k^2} -\frac{1}{2}\frac{M^4_J}{k^4} + ...
\eeq
leads to 
\beq
V^{(1)} (\phi)= \frac{1}{64 \pi^2}\left[Str \mathcal{I}  \int \frac{d^4 k}{(2 \pi)^4} \log k^2 \,  + Str \mathcal M^2(\phi)  \int \frac{d^4 k}{(2 \pi)^4}  \frac{1}{k^{2}}  \,+ ...\right] \,.
\eeq

If a UV cutoff $\Lambda$ is introduced the first term is a pure (cosmological) constant term with coefficient proportional
to $Str \mathcal{I} =n_B-n_F$ which vanishes in theories with equal number of bosonic ($n_B$) and fermionic ($n_F$) degrees of freedom (as e.g.~supersymmetric theories). 
The second term is of order $\Lambda^2$ and determines the presence of quadratic divergences 
at the one-loop level. Therefore quadratic divergences are absent provided that $Str \mathcal M^2=0$.  
More precisely one can even tolerate $Str \mathcal M^2=constant$ since this would correspond to a shift of the zero point energy which remains undetermined in the absence of coupling to gravity. 
In theories with exact or spontaneously broken supersymmetry the vanishing of $Str \mathcal M^2$ is fulfilled whenever the trace-anomaly vanishes. The soft supersymmetry breaking terms are defined as those non supersymmetric terms that can be added to the supersymmetric Lagrangian without spoiling the constancy of $Str \mathcal M^2$. 

It is important to stress ~\cite{Einhorn:1992um} that $Str \mathcal M^2$ is a function of $\phi$. 
In a supersymmetric theory $Str \mathcal M^2=0$
is true for any value of $\phi$, and thus represents simultaneous satisfaction of three sets of constraints, corresponding to terms which are proportional to $\phi^0$, $\phi$ and $\phi^2$, respectively.  
In a generic theory the vanishing of $Str \mathcal M^2$  will instead occur only for specific values of $\phi$.

Consider now the SM with Higgs potential 
\beq
V(\phi)= -\frac{m^2}{2} \phi^2 + \frac{\lambda}{4} \phi^4
\eeq
%
In fact $Str \mathcal M^2(\phi)$ can be seen as a function of the renormalization scale $\mu \sim \phi$,
as given by
\begin{equation}
Str \mathcal M^2(\phi) = H (\mu)+ 3 G(\mu) + 6 W(\mu) + 3 Z(\mu) -12 T(\mu)
\label{eq:strSM}
\end{equation}
where the numerical coefficients in Eq.~(\ref{eq:strSM}) come from the number of degrees of freedom of the physical Higgs (one), the Goldstone bosons (three), the massive gauge bosons $Z$ (three) and $W$ (six) and the top, a Dirac fermion (twelve). In fact
\begin{align}
H(\mu)&= -m^2(\mu) +3  \lambda(\mu) \phi^2 \nonumber\\
G(\mu)&= -m^2 (\mu)+  \lambda(\mu) \phi^2 \nonumber\\
W(\mu)&= \frac{1}{4} g^2(\mu) \phi^2 \nonumber\\ 
Z(\mu)&=  \frac{1}{4} (g^2(\mu)+g'^2(\mu)) \phi^2\nonumber\\ 
T(\mu)&= \frac{1}{2} y_t^2(\mu) \phi^2 \,,
\end{align}
$y_t$ being the top Yukawa coupling and $g,g'$ the electroweak gauge couplings.
When $\phi=v\approx 246$ GeV, we have that $H,W,Z,T$ become the physical masses $m_H^2,m_W^2,m_Z^2,m_t^2$ while $G=0$. 
Note that there is no term linear in $\phi$ in (\ref{eq:strSM}), because the SM does not have a cubic scalar
invariant term in the Lagrangian.  
Clearly in the SM it is not possible to have  $Str \mathcal M^2=0$ for general $\phi$,
since the $m^2$ terms in  (\ref{eq:strSM}) do not cancel. The vanishing of $Str \mathcal M^2$ will happen
only at some specific value of $\phi$. Since in the RGE we are identifying $\phi$ with the renormalization scale $\mu$,
for large field values the terms proportional to $\phi^2$ in (\ref{eq:strSM}) will neatly dominate;
in other words the absence of quadratic divergences is provided by the condition~\cite{Einhorn:1992um}
\beq
\frac{\partial Str \mathcal M^2(\phi)}{\partial \phi^2}=6 \lambda(\mu)  + \frac{9}{4} g^2(\mu)  + \frac{3}{4} g'^2(\mu) - 6  y_t^2(\mu)=0\,,
\label{VCbis}
\eeq
that is precisely the Veltman condition at 1-loop since the right hand side of Eq.~(\ref{VC1}) can be written in terms of running couplings
as
  \beq
C_{V1}(\mu) =6 \lambda(\mu)  + \frac{9}{4} g^2(\mu)  + \frac{3}{4} g'^2(\mu) - 6  y_t^2(\mu)\,.
\label{CV1SM}
\eeq 

Notice that including two-loop (or higher-loop) corrections will modify the condition (\ref{CV1SM}) by a loop suppressed $\mathcal O[1/(4\pi)^2]$ quantity which will translate into a tiny modification of the one-loop Veltman scale $\mu_{V_1}$~\cite{Alsarhi:1991ji,Hamada:2012bp}.

%

\section{The Veltman condition in the Standard Model}
\label{VC}
\begin{figure}[htb]
\vskip .5cm 
 \begin{center}
\includegraphics[width=11cm]{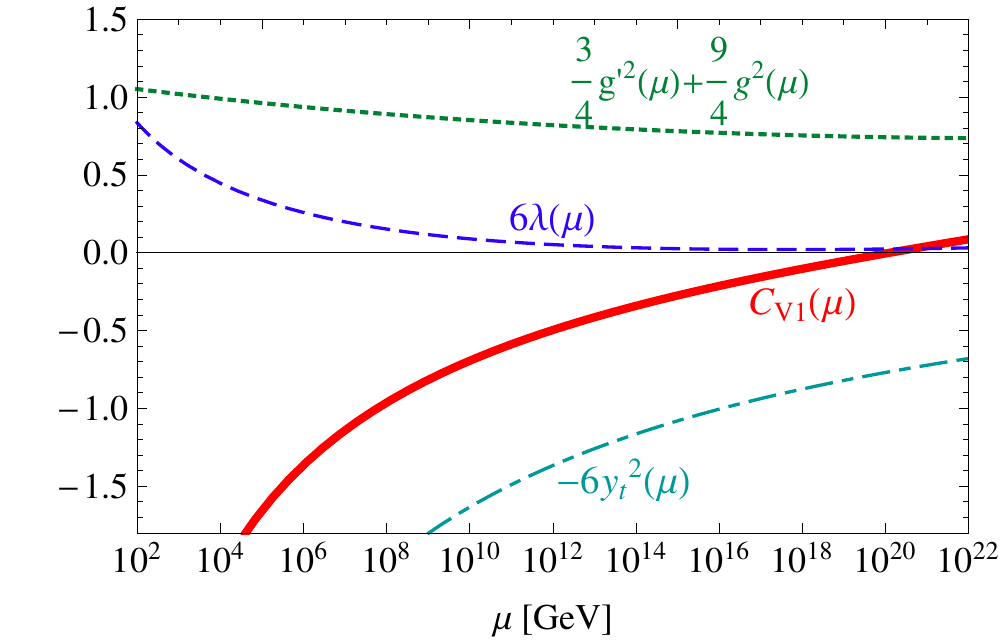}   
 \end{center}
\caption{\baselineskip=15 pt
Running of the Veltman condition $C_{V1}(\mu)$ and its various contributions as functions 
of the renormalization scale $\mu$. The plot has been obtained by choosing:
$m_H=126$ GeV for the Higgs mass,  $\overline{m_t}(m_t)=161.5$ GeV for the running top mass in the $\overline{\rm MS}$ scheme evaluated at the top mass, $\alpha_3(m_Z) =0.1196$ for the strong coupling constant evaluated at the $Z$ mass. }
\label{fig-VC1SM}
\vskip .1 cm
\end{figure}

In terms of running couplings the VC1 reads as $C_{V1}(\mu) =0$.
An example of the running of the above quantities is provided in Fig.~\ref{fig-VC1SM}. 
We have performed a NNLO running (three-loop for beta functions \cite{gauge} and two-loop for 
the matchings \cite{match}) as discussed in~\cite{Masina:2012tz}. The plot has been obtained by choosing the following values of the input parameters: $m_H=126$ GeV for the Higgs mass,  $\overline{m_t}(m_t)=161.5$ GeV for the running top mass in the $\overline{\rm MS}$ scheme evaluated at the top mass, $\alpha_3(m_Z) =0.1196$ for the strong coupling constant evaluated at the $Z$ mass. We recall that, according to SM fit results by the Particle Data Group~\cite{PDG}: $\overline{m_t}(m_t)=163.71 \pm 0.95$ GeV and $\alpha_3(m_Z)=0.1196 \pm 0.0017$ at 1$\sigma$.

In the present work we display all results as a function of $\overline{m_t}(m_t)$, as suggested in~\cite{Alekhin:2012py} and done in~\cite{Masina:2012tz}. This allows to avoid the theoretical error due to the matching between the pole mass and the running $\overline{\rm MS}$ mass of the top quark. In order to make a link with the pole top mass value $M_t$, we note that:
\textbf{i)} According to Ref.~\cite{Alekhin:2012py} the pole top mass value is just about $10.0$ GeV larger than the running top mass;
\textbf{ii)} According to the most recent analysis of Ref.~\cite{Buttazzo:2013uya}
the top pole mass mean value obtained by combining  the latest data from ATLAS, CMS and CDF is $M_t = 173.36 \pm 0.65 \pm 0.3$ GeV at $1\sigma$, where the last uncertainty 
is of theoretical origin and it is due to non-perturbative effects of order $\Lambda_{QCD}$.
Here we consider it is safe to take the experimental range of the running top mass to be given by $\overline{m_t}(m_t)=163.5 \pm 2.0$ GeV at $2\sigma$. The value $\overline{m_t}(m_t)=161.5$ GeV selected in Fig.~\ref{fig-VC1SM} is thus the lowest acceptable one 
at $2\sigma$. Notice that the contribution to the VC1 from the Higgs quartic coupling (dashed line) is smaller than the gauge (dotted line) and Yukawa (dot-dashed line) contributions. Increasing the top mass the Higgs quartic coupling at high scales decreases and, as a consequence, the scale $\mu$ where the VC1 is fulfilled increases. 
This means that in the SM the VC1 is typically realized at too high a scale, slightly larger than the Planck scale. 

\begin{figure}[htb]
\vskip .5cm 
 \begin{center}
\includegraphics[width=11.3cm]{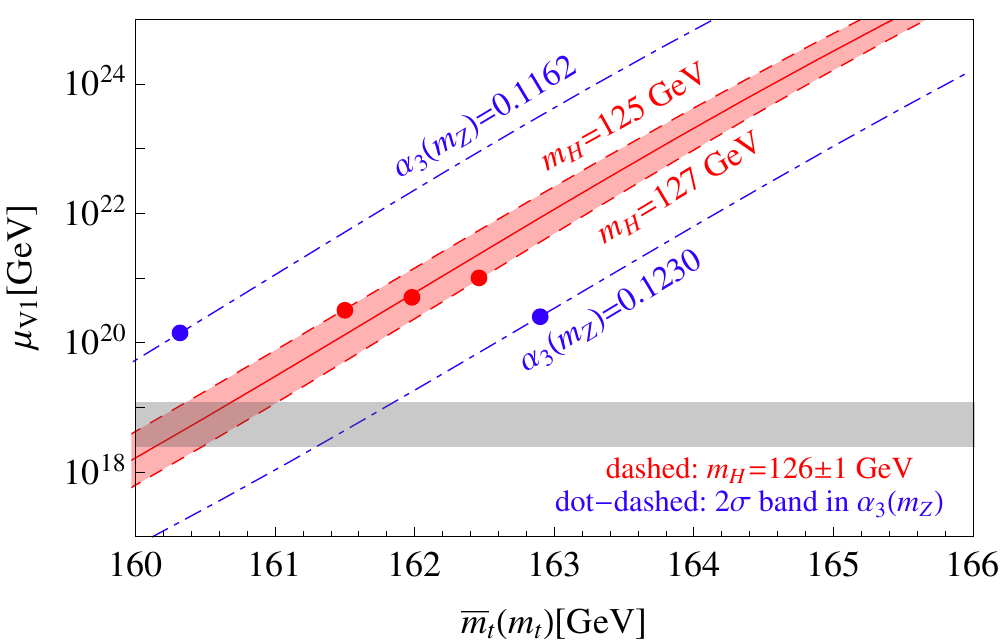} 
 \end{center}
\caption{\baselineskip=15 pt
The dependence of $\mu_{V_1}$ on the running top mass $\overline{m_t}(m_t)$. The solid line corresponds to $m_H=126$ GeV and  $\alpha_3(m_Z) =0.1196$.  The shaded band is obtained by varying $m_H$ by $\pm 1$ GeV. The dot-dashed lines are obtained by varying $\alpha_3(m_Z)$ in its $2\sigma$ range~\cite{PDG}. The shaded grey region emphasizes the range between the Planck mass ($M_{Pl}=1.2\times 10^{19}$ GeV)
and the  reduced Planck mass ($M_{Pl}/\sqrt{8\pi}$).  }
\label{fig-muV1SM}
\vskip .1 cm
\end{figure}

This is better illustrated in Fig.~\ref{fig-muV1SM}, where we plot the scale  $\mu_{V_1}$ such that $C_{V1}(\mu_{V_1})=0$, as a function of the running top mass in the $\overline{\rm MS}$ scheme. The solid line is obtained for $m_H=126$ GeV and $\alpha_3(m_Z) =0.1196$. The shaded region between the dashed lines is obtained by keeping $\alpha_3(m_Z)$ fixed and varying $m_H$ in the range $126 \pm 1$ GeV.  
The dot-dashed lines are obtained by keeping $m_H=126$ GeV and varying $\alpha_3(m_Z)$ in its $2\sigma$ range~\cite{PDG}. The dots on the lines signal the value of $\overline{m_t}(m_t)$ such that a second vacuum degenerate with the electroweak one appears. For instance focussing on the case $m_H=126$ GeV and  $\alpha_3(m_Z) =0.1196$ (solid line), this happens for $\overline{m_t}(m_t) \simeq 162$ GeV, and the second degenerate minimum turns out to be located at $\mu=4.3 \times 10^{17}$ GeV~\footnote{Note that the scale at which this happens does not coincide with the scale at which the VC1 is satisfied in Fig.~\ref{fig-muV1SM}.}. For larger (smaller) values of $\overline{m_t}(m_t)$ the Higgs potential is thus stable (metastable).

One realizes that the uncertainty on $\mu_{V_1}$ due to the $2\sigma$ range of $\alpha_3(m_Z)$ is larger than the uncertainty due to the range of the Higgs mass.
Anyway there is little room in the experimentally allowed SM parameter space for the VC1 to be satisfied at sub-Planckian scales: this happens only for the highest possible values of the Higgs mass and $\alpha_3(m_Z)$ and for the lowest possible values of $\overline{m_t}(m_t)$. 

\section{SM merging into high-scale MSSM}
\label{MSSM}

As we pointed out in section~\ref{introduction} one can interpret the realization of the Veltman condition around the Planck scale as an indication that we have a fine-tuned Standard Model up to the scale $ \mu_{V_1}$ which merges into a version of the MSSM which is in turn possibly embedded into a more fundamental theory, as e.g.~a superstring theory. This idea of a high-scale MSSM~\footnote{Notice the difference between our construction, i.e.~high-scale supersymmetry where the effective theory below $\mu_V$ is just the Standard Model, and the so-called split supersymmetry, where Higgsino and gaugino masses are protected by some symmetry and remain in the low energy theory.} has been put forward recently using ideas based on the stability 
condition~\cite{Hall:2009nd,Hebecker:2012qp,Ibanez:2013gf}. We will show here that a similar idea can be 
implemented using as an input the Veltman condition~\footnote{For an early study previous to the Higgs discovery see~\cite{Casas:2004gh}.}.

As the Standard Model is valid to some high-scale it is fine-tuned. At this point we should give up with solving the SM hierarchy (along with the cosmological constant) problem unless a symmetry reason in the UV merging theory, or some kind of environmental selection, is provided as it would be the case of the huge number (landscape) of string vacua. In any way if we denote by $H_1$ and $H_2$ the MSSM Higgs doublets giving mass to the down and up quarks, respectively, with a superpotential $W=\mu \mathcal H_1\cdot \mathcal H_2$, the quadratic terms in the tree-level potential are
\beq
V_2=m_1^2 |H_1|^2+m_2^2|H_2|^2+m_3^2(H_1\cdot H_2+h.c.)
\eeq
where $H_1\cdot H_2=H_1^a \varepsilon_{ab} H_2^b$ ($\varepsilon_{12}=-1$) and the soft breaking terms $m^2_{1,2,3}\simeq \mu_{V_1}^2$. The fine-tuning required to have a single light Higgs is provided by the condition 
\beq
 m_3^4 \simeq m_1^2 m_2^2
\label{FT1}
\eeq
which should be satisfied up to $\mathcal O[m_H^2(m_1^2+m_2^2)]$, and the light and heavy eigenstates (i.e.~the SM Higgs $H$ and its orthogonal combination $H_{Heavy}$) are
\begin{align}
H&=\cos\beta H_1-\sin\beta \widetilde H_2\nonumber\\
H_{Heavy}&=\sin\beta H_1+\cos\beta \widetilde H_2
\label{combinaciones}
\end{align}
where 
\beq
\tan\beta\simeq |m_1|/|m_2|\,,
\eeq
$\widetilde H_2\equiv \varepsilon H_2^*$ and the mass of the heavy Higgs is $m_{Heavy}^2=m_1^2+m_2^2$. 
The projection over the SM-Higgs 
\beq
H_1\cdot H_2=-\sin\beta\cos\beta |H|^2,\quad |H_1|^2+|H_2|^2=|H|^2
\label{projection}
\eeq
yields the light state $H_{SM}$. At this point we are not going to argue about the origin of this fine-tuning but we will just accept that the SM merges below the scale $\mu_{V_1}$. This means that we are assuming that all soft breaking parameters (soft-breaking masses, gaugino and Higgsino masses and soft trilinear couplings) are $\mathcal O(\mu_{V_1})$ so that we can match the MSSM and the SM at the scale $\mu_{V_1}$.

\begin{figure}[htb]
\vskip .5cm 
 \begin{center}
 \includegraphics[width=7.45cm]{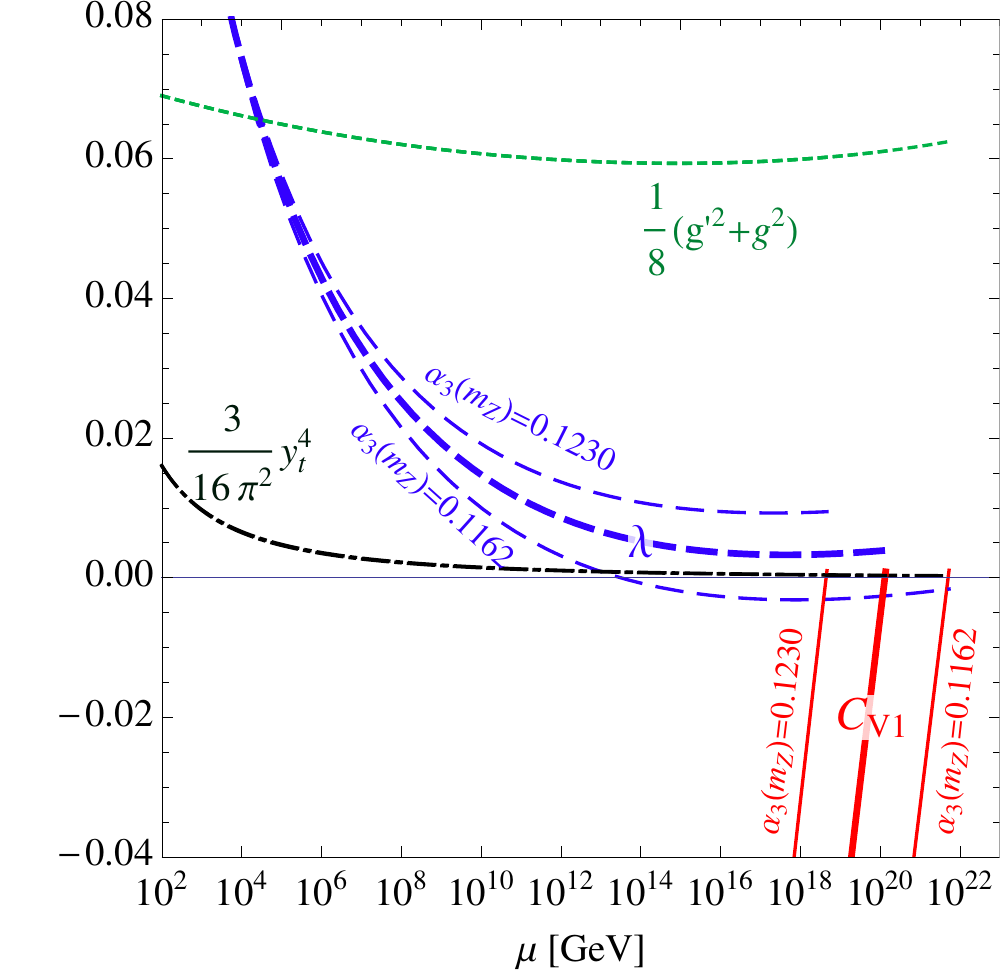}     \, \, \, \, \includegraphics[width=7.65cm]{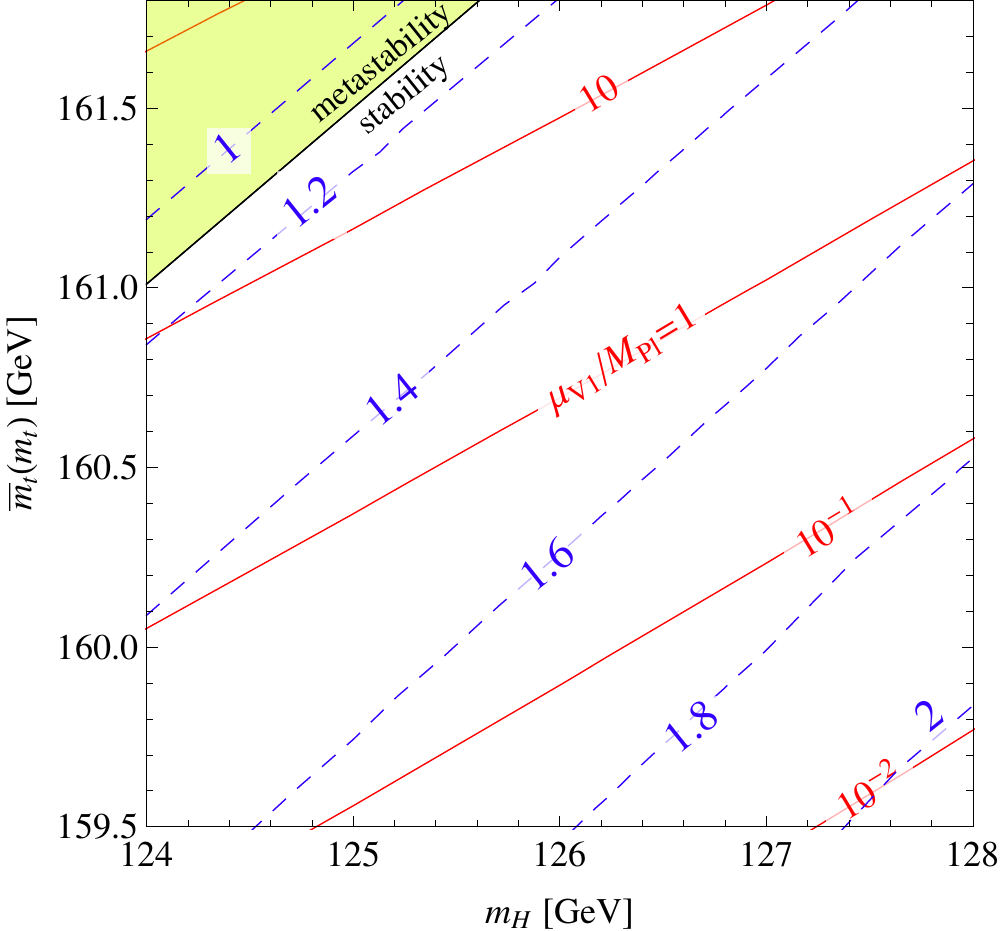}   
  \end{center}
\caption{\baselineskip=15 pt
Left panel: Example of merging of SM with the MSSM. The plot has been obtained by choosing $m_H=126$ GeV,  $\overline{m_t}(m_t)=161.5$ GeV. The thicker dashed and solid lines are obtained by choosing $\alpha_3(m_Z) =0.1196$; the thinner dashed and solid lines show the effect of varying $\alpha_3(m_Z)$ by $2\sigma$ (this has a negligible impact on the running of the gauge and top Yukawa couplings).
Right panel: value of $\mu_{V_1}/M_{Pl} $ (solid red) and upper value on $\tan\beta$ (dashed blue). We fixed $\alpha_3(m_Z) =0.1196$.}
\label{fig-VC1SMsusy}
\vskip .1 cm
\end{figure}
Recall that the merging with the MSSM at $\mu_{V_1}$ requires  
\beq
\lambda(\mu_{V_1})= \frac{1}{8} \left[g'^2(\mu_{V_1})+g^2(\mu_{V_1}) \right] \cos^2 2\beta + \frac{3}{16\pi^2}y_t^4(\mu_{V_1})x_t^2\left(1-\frac{x_t^2}{12}\right) \, , 
\label{matchMSSM}
\eeq
where the last term is a threshold correction coming from a possible mixing in the stop sector, $x_t=(A_t-\mu/\tan\beta)/m_S$ and we are identifying the common supersymmetric mass $m_S\simeq \mu_{V_1}$. In Fig.~\ref{fig-VC1SMsusy} we show the running of $C_{V1}(\mu)$ for particular values of the top and Higgs masses, as well as the running of the different terms in Eq.~(\ref{matchMSSM}). In particular we can see that the running of the threshold correction $3 y_t^4/16\pi^2$ brings it to extremely small values at $\mu_{V_1}$ so that we can, for all purposes, neglect this correction and thus we will assume from here on the case of no mixing, i.e.~$x_t=0$. 
This  means in particular that 
\beq
 0 \le \lambda(\mu_{V_1})\le  \frac{1}{8} \left[g'^2(\mu_{V_1})+g^2(\mu_{V_1}) \right]\, ,
\eeq
where equality with the left hand side (right hand side) holds for $\tan \beta=1$ ($\tan \beta \gg 1$).
For instance for the input values adopted in the left panel of Fig.~\ref{fig-VC1SMsusy}  
one realizes that it is possible to achieve a viable merging with the MSSM for $\alpha_3(m_Z) \gtrsim 0.1180$. However for a large enough top mass such that $\lambda$ is negative at $\mu_{V_1}$, the merging with the MSSM is not possible for any allowed value of $\alpha_3(m_Z)$.

We can also determine the value of $\tan \beta$ for each point in the plane
$[m_H,\overline{m_t}(m_t)]$. Such value becomes an upper bound on $\tan \beta$ as we are neglecting the mixing and we are assuming $x_t\simeq \mathcal O(1)$ such that the threshold correction to the quartic coupling is positive.
As an example, we consider for definiteness $\alpha_3(m_Z) = 0.1196$ and display in the right panel of Fig.~\ref{fig-VC1SMsusy}
the isocurves of constant $\mu_{V_1}/M_{Pl} $ (solid red) and upper value on $\tan\beta$ (dashed blue). We see that a large $\tan\beta$ is not possible for the Veltman condition to be satisfied below the Planck scale: in fact we are able to fix the upper bound on $\tan\beta$ as $\tan\beta\lesssim 2$. 
As we can see from the plots in Fig.~\ref{fig-VC1SMsusy} from the Veltman condition the merging of the SM at low energy into the MSSM at high energy can only be done, for realistic Higgs and top quark masses, at slightly trans-Planckian scales which sheds doubts on the consistency of the whole procedure as gravitational effects are never considered. 

Moreover on purely experimental grounds we do not expect the Standard Model to be the effective theory at energies presently explored by LHC as there is no valuable candidate to Dark Matter. From that point of view some SM extension should be favored as a candidate to low-energy effective theory. A quick glance at Fig.~\ref{fig-VC1SM} shows that the (negative) contribution of the top quark Yukawa coupling in the Veltman condition is responsible for its fulfillment at very high scales. So adding fermions to the SM (e.g.~vector like fermions) coupled to the Higgs field will only worsen the situation. It is clear that we need some bosonic contribution to lower the Veltman scale $\mu_V$ and from that point of view one can envisage a generic situation with bosonic Dark Matter. In the next section we will consider the simplest such model where a singlet is coupled to the SM Higgs, a model already considered in the literature for various purposes~\cite{NSM}, including the improvement of fine-tuning for the Higgs field in the Standard Model~\cite{Kundu:1994bs,Grzadkowski:2009mj,Chakraborty:2012rb}. We will consider a complex singlet, instead of a real one, to make it simpler the merging with the MSSM extended by a singlet superfield, a supersymmetric theory which has also been extensively studied in the literature~\cite{Ellwanger:2009dp} and is usually dubbed as the NMSSM~\footnote{The name NMSSM is usually reserved to the MSSM plus a singlet chiral superfield, when the scalar component of the singlet acquire a non-zero VEV which makes it possible to give a technical solution to the supersymmetric $\mu$-problem. Independently on whether our singlet does or does not not acquire any VEV, for simplicity and notational convenience we will occasionally keep on calling the model NMSSM by an abuse of language.}.

\section{SM extension with singlets}
\label{singlets}
Let us now consider the simplest SM extension with a complex singlet $S$, with a general  potential given by
\beq
V=V_{SM}+\lambda_S|S|^4+2\lambda_{SH} |H|^2|S|^2+M_S^2|S|^2+\cdots
\label{quarticS}
\eeq
where 
\beq
V_{SM}=-m^2|H|^2+\lambda|H|^4\ .
\eeq
We will consider the particular case $\lambda_S(\mu_{V_1})=0$ which will have a simple merging with the NMSSM as we will see in the next section~\footnote{Of course even if we consider $\lambda_S=0$ at the merging scale it will be generated at lower scales by the renormalization group equations (RGE), although its impact should be tiny as we have checked numerically.}.

As the mass of the complex singlet in the presence of the background field $\phi$ from (\ref{quarticS}) is given by
\beq
m_S^2(\phi)=M_S^2(\mu)+\lambda_{SH}(\mu)\,\phi^2
\eeq
the Veltman condition for the cancellation of quadratic divergences (\ref{VCbis}) reads as
\beq
6\lambda(\mu)+2\lambda_{SH}(\mu)+\frac{9}{4}g^2(\mu)+\frac{3}{4}g'^2(\mu)-6 y_t^2(\mu)=0\,.
\label{VCsinglet}
\eeq
where the contribution of the $S$ field to $Str \mathcal M^2$ is $2 m_S^2(\phi)$ and thus the factor of two in Eq.~(\ref{VCsinglet}) is the number of degrees of freedom of the complex singlet.

The introduction of a light singlet with Lagrangian given by the potential (\ref{quarticS}) creates a \textit{second} hierarchy problem even if the scalar singlet $S$ is heavier (at the TeV scale) than the SM Higgs and independently on whether or not it does acquire a VEV~\cite{Chakraborty:2012rb}. In particular the coupling $2\lambda_{SH}|S|^2|H|^2$ creates a quadratic sensitivity on the scale for the $|S|^2$ term. This quadratic sensitivity can only be canceled by a fermion mass contribution in the presence of the background field $S$. In particular the minimal extension is done by adding to the SM the scalar $S$ \textit{and} a (singlet) Weyl fermion $f$ with a mass Lagrangian
\be
\mathcal L_{f} =-\frac{1}{2}(m_f+\lambda_f S) f^\alpha f_\alpha+h.c.\ .
\label{Lagrangianf}
\ee
and thus $M_{1/2}(S)=m_f+\lambda_f S$. The Veltman condition along the $|S|^2$-direction should be then understood as the $S$-dependent part of $Str \mathcal M^2(\phi,S)$. It
can then be written as
\beq
8\lambda_{SH}(\mu_{V_1})-2\lambda_f^2(\mu_{V_1})=0\,.
\label{VCS}
\eeq
where the first term comes from the Higgs doublet $H$ (four degrees of freedom) and the second one from the Weyl fermion $f$ (two degrees of freedom)~\footnote{As we will see in the next section there is also a quadratically divergent tadpole for the singlet $S+S^*$ from the supertrace term $M_{1/2}^\dagger(S) M_{1/2}(S)$ which should be consistently cancelled by the matching conditions of the supersymmetric merging theory.}.

The one-loop beta functions for the interactions in Eqs.~(\ref{quarticS}) and (\ref{Lagrangianf}) are given by~\cite{EliasMiro:2012ay,Lee}
\bea
(4\pi)^2 \frac{d \lambda}{dt} &=& (12 y_t^2 -3 g'^2 -9 g^2) \lambda -6 y_t^4 +\frac{3}{8} \left[2 g^4+(g'^2+g^2)^2\right] +24 \lambda^2 +4 \lambda_{SH}^2 \, , \nonumber \\
(4\pi)^2 \frac{d \lambda_{SH}}{dt} &=& \frac{1}{2}(12 y_t^2 -3 g'^2 -9 g^2) \lambda_{SH} + 
4\lambda_{SH} (3\lambda + 2 \lambda_S) +8 \lambda_{SH}^2 +4\lambda_{SH}\lambda_f^2\, ,\nonumber\\
(4\pi)^2 \frac{d \lambda_{S}}{dt} &=& 8 \lambda_{SH}^2 +20 \lambda_S^2+8\lambda_S\lambda_f^2-\lambda_f^4\, , \\ \nonumber
(4\pi)^2 \frac{d \lambda_{f}}{dt} &=& 6\lambda_f^3\, ,
\label{RGE}
\eea
where $t=\ln \mu/m_Z$.

\begin{figure}[h!]
\vskip .5cm 
 \begin{center}
\includegraphics[width=7.5cm]{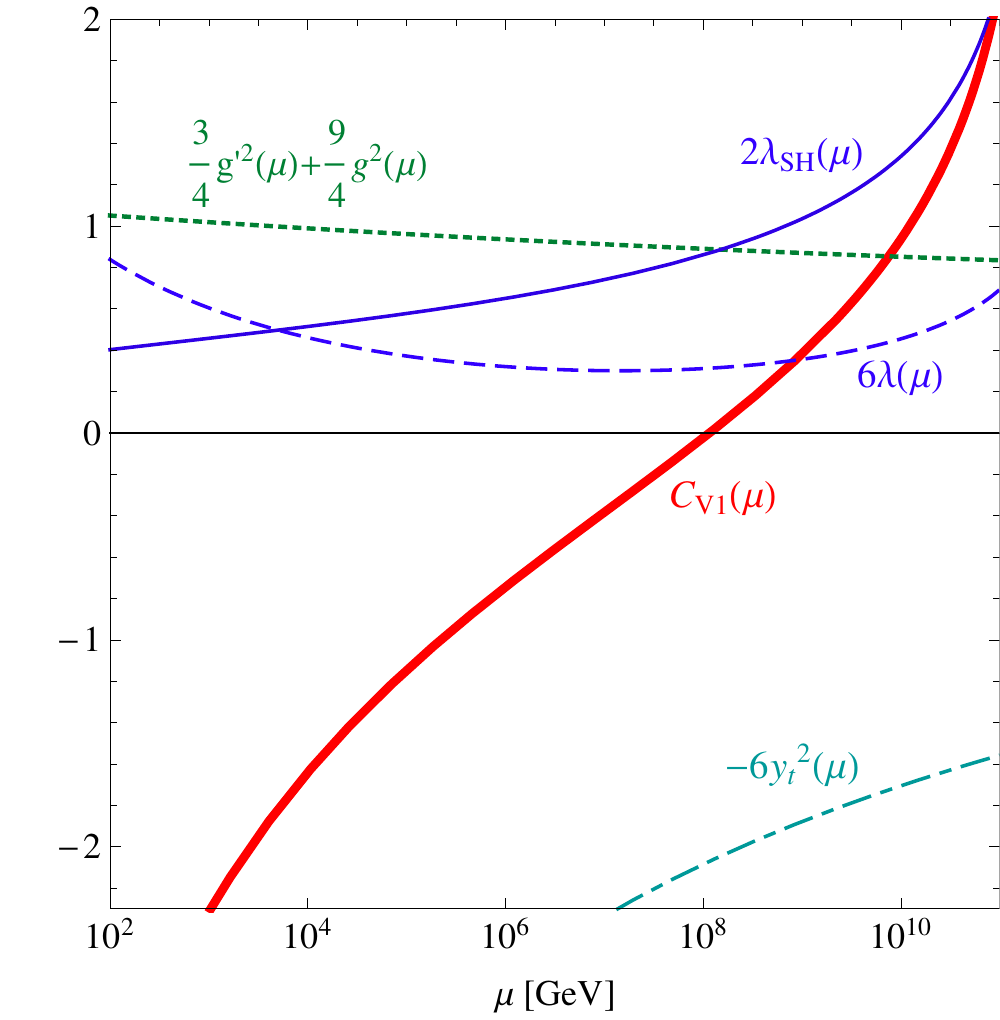}   \,\, \includegraphics[width=7.7cm]{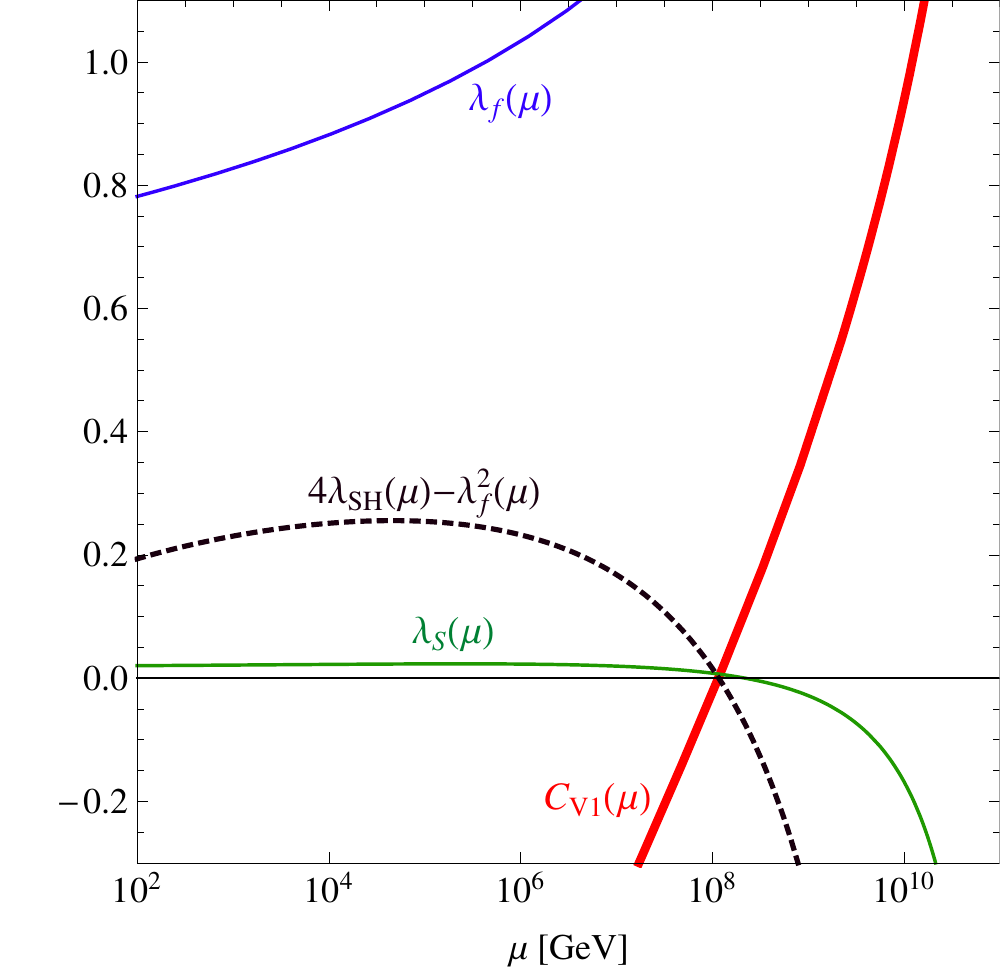}   
 \end{center}
\caption{\baselineskip=15 pt
Left panel: Same as in Fig.~\ref{fig-VC1SM} but in the model defined by the potential in Eqs.~(\ref{quarticS}) and (\ref{Lagrangianf}).  Right panel: running of the Veltman condition along the $S$ direction and of $\lambda_S$ and $\lambda_f$.  
Both panels are obtained by choosing  $\lambda_{SH}(m_Z)=0.2$, $\lambda_S(m_Z)=0.02$, $\lambda_f(m_Z)=0.78$, $m_H=126$ GeV,  $\overline{m_t}(m_t)=163.5$ GeV, $\alpha_3(m_Z) =0.1196$.}
\label{fig-VC1SMS}
\vskip .1 cm
\end{figure}

We plot in the left panel of Fig.~\ref{fig-VC1SMS} the new Higgs VC1 (\ref{VCsinglet}) and the various terms contributing to it. 
%
%
There are two main differences with respect to the SM case shown in Fig.~\ref{fig-VC1SM}, both of them contributing to lowering the scale at which VC1 is satisfied, $\mu_{V_1}\simeq 10^8$ GeV:
\begin{itemize}
\item
The first difference is that the function $\lambda(\mu)$, the SM quartic coupling, takes larger values than in the pure SM case. The origin of this increase is the contribution of the term  $4 \lambda_{SH}^2$ in its RGE, the first equation in (\ref{RGE}). The largish value of $\lambda(\mu)$ tends then to compensate the negative contribution of $-6y_t^2(\mu)$.
\item
The second difference is the very existence of the term $2\lambda_{SH}$ in the VC1 (\ref{VCsinglet}) whose value is sizeable as can be seen from Fig.~\ref{fig-VC1SMS} and tends then to compensate the negative contribution of the top quark Yukawa coupling.
\end{itemize}
 In the right panel of Fig.~\ref{fig-VC1SMS} we also show the VC1 along the $S$-direction, Eq.~(\ref{VCS}), as well as the running of couplings $\lambda_S(\mu)$ and $\lambda_f(\mu)$ whose boundary conditions are fixed at the scale $\mu_{V_1}$ by $\lambda_S(\mu_{V_1})=0$ and Eq.~(\ref{VCS}). The corresponding initial values of the couplings are indicated in the caption of Fig.~\ref{fig-VC1SMS}.
    
\begin{figure}[h!]
\vskip .5cm 
 \begin{center}
\includegraphics[width=10.cm]{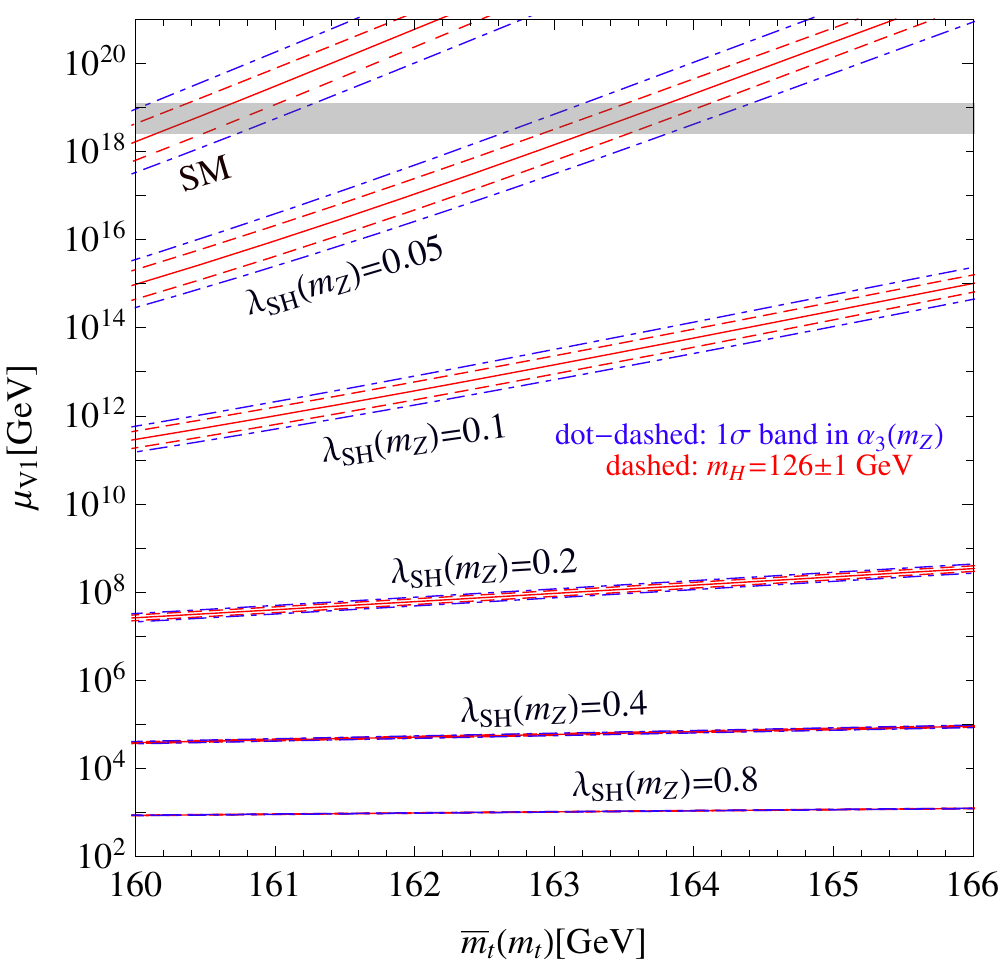} 
 \end{center}
\caption{\baselineskip=15 pt
The dependence of $\mu_{V_1}$ on the running top mass for various values of $\lambda_{SH}(m_Z)$. The solid lines are obtained for $m_H=126$ GeV and $\alpha_3(m_Z) =0.1196$.
The dashed (red) lines show the effect of varying $m_H$ in the range $126 \pm 1$ GeV; 
the dot-dashed (blue) lines show the effect of varying $\alpha_3(m_Z) $ in its $1\sigma$ range, namely 
$\alpha_3(m_Z) =0.1196 \pm 0.0017$.  
}
\label{fig-muV1SMS}
\vskip .1 cm
\end{figure}

 As we can see from Fig.~\ref{fig-VC1SMS} the reduction in the value of $\mu_{V_1}$ is sizable even for a small value of $\lambda_{SH}$. In fact the larger $\lambda_{SH}(m_Z)$ the stronger the reduction of the $\mu_{V_1}$ scale. The obvious question is how much can we lower the $\mu_{V_1}$ scale and in particular whether or not it can be pushed towards the TeV scale. To answer this question we have plotted in Fig.~\ref{fig-muV1SMS} the scale $\mu_{V_1}$ as a function of the running top mass for various values of $\lambda_{SH}(m_Z)$. For each curve we have selected the values of $\lambda_{S}(m_Z) $ and $\lambda_{f}(m_Z)$ so that both Veltman conditions, Eqs.~(\ref{VCsinglet}) and (\ref{VCS}), are satisfied along with $\lambda_S(\mu_{V1})=0$. 
In particular for $\lambda_{SH}(m_Z) =(0.05,0.1,0.2,0.4,0.8)$ we have chosen respectively $\lambda_{S}(m_Z) \approx(0.003,0.007,0.02,0.03,0.05)$ and $\lambda_{f}(m_Z) \approx(0.41,0.56,0.78,1.1,1.59)$.
We have also shown in Fig.~\ref{fig-muV1SMS}, for the sake of comparison, the SM case (where $\lambda_{SH}=0$) which clearly shows, as it was pointed out in previous sections, that identifying $\mu_{V_1}$ with the Planck scale requires borderline values of the top quark and Higgs masses and of $\alpha_3(m_Z)$. Moreover Fig.~\ref{fig-muV1SMS} shows that, for fixed values of $\overline m_t(m_t)$, $m_H$ and $\alpha_3(m_Z)$, the scale $\mu_{V_1}$ exhibits an exponential sensitivity to the value of $\lambda_{SH}(m_Z)$ and for $\lambda_{SH}(m_Z)\lesssim \mathcal O(1)$ values of $\mu_{V_1}\gtrsim$ TeV, where the theory is perturbative, can be reached. Clearly the UV completion of the theory merging at the scale $\mu_{V_1}$, should most probably become non- perturbative at some scale below $M_{Pl}$ when the coupling in the UV theory matching $\lambda_{SH}(\mu_{V_1})$ reaches a Landau pole. 

In the next section we will study the merging with the simplest supersymmetric theory which consists in the MSSM plus the addition of singlet chiral superfields. In particular we will see that adding to the NMSSM an additional chiral singlet $\mathcal T$, with coupling and masses satisfying some relations, condition (\ref{VCS}) can be satisfied.

\section{Merging to the MSSM with singlets}
\label{NMSSM}

We will assume here that the actual model, the SM plus the complex singlet $S$, merges into a supersymmetric theory. As far as the Veltman condition along the SM Higgs is concerned  the simplest such theory is the MSSM plus a chiral supersymmetric singlet $\mathcal S$ which we are calling NMSSM with a superpotential $W$ given by
\begin{equation}
W=\lambda_2 \mathcal S\mathcal H_1\cdot \mathcal H_2+
\mu_H \mathcal H_1\cdot \mathcal H_2+\frac{1}{2}\mu_S \mathcal S^2
\label{cubic}
\end{equation}
and the potential for the Higgs and singlet sectors, with the addition of soft breaking terms (including a trilinear soft coupling $A_\lambda$) is written as
\begin{align}
V&=m_1^2 |H_1|^2+m_2^2|H_2|^2+m_3^2(H_1 \cdot H_2+h.c.)+M_S^2|S|^2\nonumber\\
&+\lambda_2^2|H_1\cdot H_2|^2+\lambda_2^2|S|^2(|H_1|^2+|H_2|^2)+\frac{g^2+g'^2}{8}(|H_1|^2-|H_2|^2)^2\nonumber\\
&+\lambda_2\mu_H S(|H_1|^2+|H_2|^2)+\lambda_2 \mu_S S^*(H_1\cdot H_2)+\lambda_2 A_\lambda SH_1\cdot H_2+h.c.\,.
\label{pot1}
\end{align}

By making the fine-tuning (\ref{FT1}) required to have a light SM Higgs $H$, and consequently the projections (\ref{combinaciones})-(\ref{projection}), we can write the potential (\ref{pot1}) as
\beq
V=V_{SM}+M_S^2|S|^2+\lambda_2^2|S|^2|H|^2+\lambda_2\left(\mu_H-\frac{\mu_S+A_\lambda}{2}\sin 2\beta\right) (S+S^*)|H|^2
\label{potencialampliado}
\eeq
with
\beq
\lambda= \frac{1}{8}(g^2+g'^2) \cos^2 2\beta+\frac{1}{4}\lambda_{2}^2\sin^2 2\beta
+\frac{3}{16\pi^2}h_t^4x_t^2\left(1-\frac{x_t^2}{12 }\right)
\label{matching}
\eeq
where we also have introduced the threshold corrections generated by integrating out the heavy stops $\tilde Q$ and $\tilde U^c$ with $x_t=(A_t-\mu/\tan\beta)/M_S$ and where we are identifying the common supersymmetric mass $M_S\simeq \mu_{V_1}$.
Then the matching with the potential (\ref{quarticS}) is done by the condition
\beq
\lambda_2^2(\mu_{V_1})=2\lambda_{SH}(\mu_{V_1})
\eeq
and the additional fine-tuning [on top of the MSSM one (\ref{FT1})] at the merging scale $\mu_{V_1}$
\beq
\tilde\mu\equiv\mu_H-\frac{\mu_S+A_\lambda}{2}\sin 2\beta=\mathcal O(\textrm{TeV})\,.
\label{FT2}
\eeq
Notice that unless we tune exactly $\tilde\mu=0$ the potential (\ref{potencialampliado}) differs from the potential (\ref{quarticS}) by the last term. In fact the last term of potential (\ref{potencialampliado}) generates a mixing between the $\mathcal R$ field, where 
\beq
S=\frac{\mathcal R+i\mathcal I}{\sqrt{2}} \, ,
\eeq
and the physical Higgs $h$ through the Lagrangian term $\propto \tilde\mu\, v\, \mathcal R(x) h(x)$. 
However the latter does not modify the conditions for the vanishing of the Veltman relations, either along the $H$ and the $S$ fields worked out in section~\ref{singlets}, as it appears in an off-diagonal entry of the squared mass matrix which obviously does not contribute to the quantity $Str \mathcal M^2(H,S)$. However as we will see the term $\propto \tilde\mu \mathcal R |H|^2$ is a necessary ingredient in the diagonal Higgs mass so as to cancel the tadpole along the $\mathcal R$ field.

In order to also fulfill the Veltman condition along the $S$ field the simplest possibility is adding to the NMSSM an extra singlet superfield $\mathcal T$ such that $\mathcal S$ is coupled to it in the superpotential $W_\mathcal T$ as
\beq
W_\mathcal T=\frac{1}{2} \lambda_3 \mathcal S \mathcal T^{\,2}+\frac{1}{2}\mu_T \mathcal T^{\,2}\,.
\label{superT}
\eeq

The scalar component of $\mathcal T$ ($T$)  is integrated out at the scale $\mu_{V_1}$ by a soft breaking mass term $m_T^2|T|^2$ where $m_T\simeq\mu_{V_1}$, it disappears from the low-energy effective theory and therefore it does not require any extra Veltman condition. The fermionic component of $\mathcal T$ denoted by $f$ has a mass from the superpotential (\ref{superT}) as
\beq
\mathcal L_{f}=-\frac{1}{2}(\mu_T+\lambda_3 S)f^\alpha f_\alpha+h.c.
\label{Lagrangianfermion}
\eeq
where we are assuming the supersymmetric parameter $\mu_T=\mathcal O$(TeV) and from the Lagrangian (\ref{Lagrangianf}) we can identify, at the matching scale $\mu_{V_1}$,
\beq
\lambda_f(\mu_{V_1})=\lambda_3(\mu_{V_1}),\quad m_f(\mu_{V_1})=\mu_T(\mu_{V_1})\,.
\label{masafermion}
\eeq

Now to cancel all quadratic divergences along the $S$ field direction we need to evaluate the supertrace in the background field $S$. From the potential (\ref{potencialampliado}) and the fermion Lagrangian (\ref{masafermion}) we can write
\beq
Str \mathcal M^2(S)=4\left[\lambda_2^2|S|^2+\lambda_2\tilde\mu(S+S^*) \right]-2\left[\lambda_3^2 |S|^2+\lambda_3\mu_T(S+S^* )\right]+\cdots
\label{strazaS}
\eeq 
where the first term represents the contribution from the Higgs $H$ in $Tr M_0^2$ (four degrees of freedom), the second term the contribution from the Weyl fermion $f$ in $Tr M_{1/2}^\dagger M_{1/2}$ (two degrees of freedom) and the ellipsis indicates $S$-independent terms. From Eq.~(\ref{strazaS}) we obtain the absence of quadratic divergences in $|S|^2$ mass terms if
\beq
2\lambda_2^2(\mu_{V_1})= \lambda_3^2(\mu_{V_1})
\eeq
and cancelation of the quadratically divergent tadpole for
\beq
\tilde\mu(\mu_{V_1})=\frac{\mu_T(\mu_{V_1})}{\sqrt{2}}\,.
\eeq
Finally the low energy effective theory is defined by the potential (\ref{potencialampliado}) and the fermion mass and Yukawa Lagrangian (\ref{Lagrangianfermion}).

\begin{figure}[htb]
\vskip .5cm 
 \begin{center}
 \includegraphics[width=8.cm]{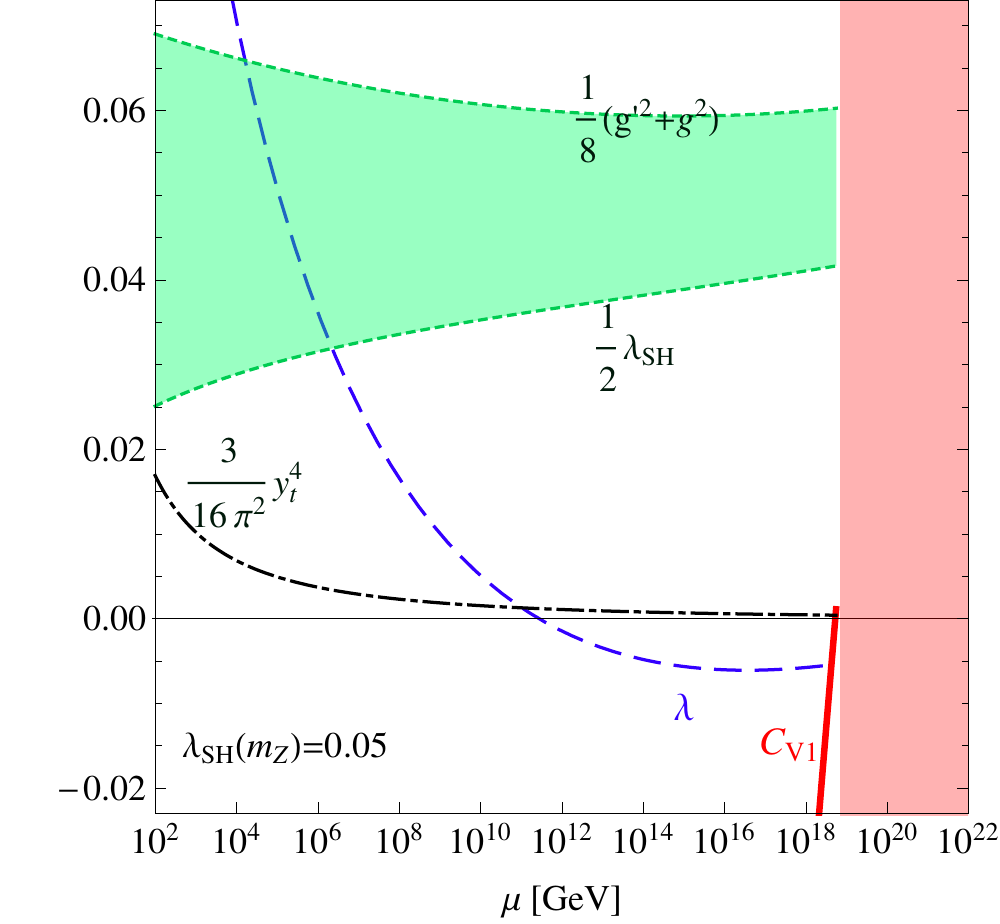}     \,\,\,\,
 \includegraphics[width=7.5cm]{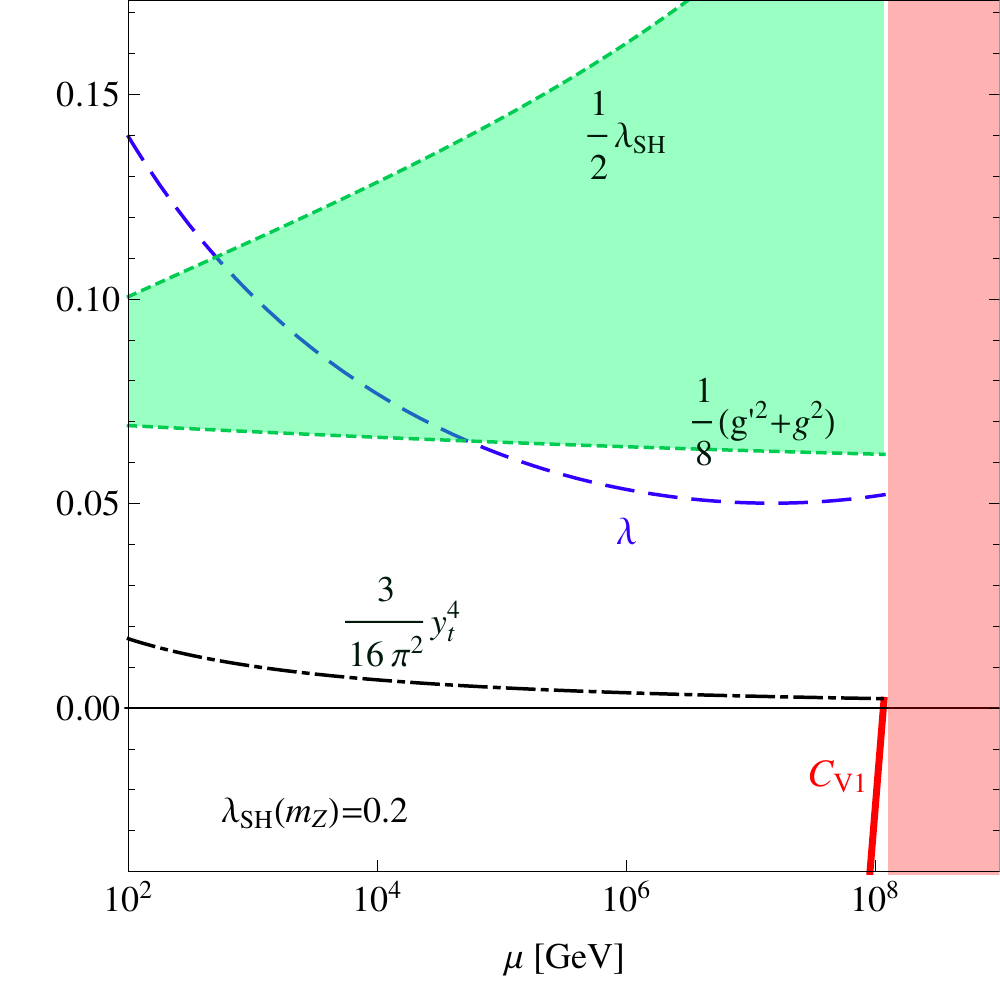}  
 \end{center}
\caption{\baselineskip=15 pt
Two examples of cases where the NMSSM merging is not viable.
The corresponding value of $\lambda_{SH}(m_Z)$ is indicated for each panel. 
For both plots we fixed  $m_H=126$ GeV, $\alpha_3(m_Z) =0.1196$, $\overline{m_t}(m_t)=163.5$ GeV.}
\label{fig-VC1SMSnonsusy}
\vskip .1 cm
\end{figure}
All parameters in the matching condition (\ref{matching}) should be considered at the scale $\mu_{V_1}$ and thus there is no guarantee that the supersymmetric theory could be matched at that scale for realistic values of the top quark and Higgs masses and an arbitrary value of $\lambda_{SH}$. The reason can be understood as follows. Neglecting the mixing in the stop sector~\footnote{For values of $\lambda_{SH}\lesssim\mathcal O(1)$ the threshold corrections provide a tiny contribution to the Higgs quartic coupling because, as we will see next, the top Yukawa coupling is driven to small values at high energy scales. Therefore from here on we will consider the case of zero mixing.} we can write the relations
\beq
{\rm Min}\left[ \frac{1}{2} \lambda_{SH} , \frac{1}{8}(g^2+g'^2) \right]  \le \lambda   \le 
{\rm Max}\left[ \frac{1}{2} \lambda_{SH} , \frac{1}{8}(g^2+g'^2) \right]  
\label{interval}
\eeq
so in the cases where $\lambda(\mu_{V_1})$ is outside the above range the merging into the NMSSM is not 
viable. This happens for small enough values of $\lambda_{SH}(m_Z)$
as it is shown in Fig.~\ref{fig-VC1SMSnonsusy} where we plot the Veltman condition for $\lambda_{SH}=0.05$ (left panel) and $\lambda_{SH}=0.2$ (right panel). In each case we have selected the values of $\lambda_{S}(m_Z) $ and $\lambda_{f}(m_Z)$ 
so that the Veltman condition in the $S$ direction, Eq.~(\ref{VCS}), is also satisfied, along with $\lambda_S(\mu_{V1})=0$. 
In particular we have chosen for the plot in the left panel $\lambda_S(m_Z) \approx0.003$ and $\lambda_f(m_Z) \approx0.41$ and for the plot in the right panel $\lambda_S(m_Z)\approx0.02$ and $\lambda_f(m_Z)\approx0.78$. As one can see the value of $\lambda(\mu_{V_1})$ is outside the interval (\ref{interval}) and any reasonable mixing with $x_t=\mathcal O(1)$ could not cope with it.

\begin{figure}[h!]
\vskip .5cm 
 \begin{center} 
   \includegraphics[width=7.8cm]{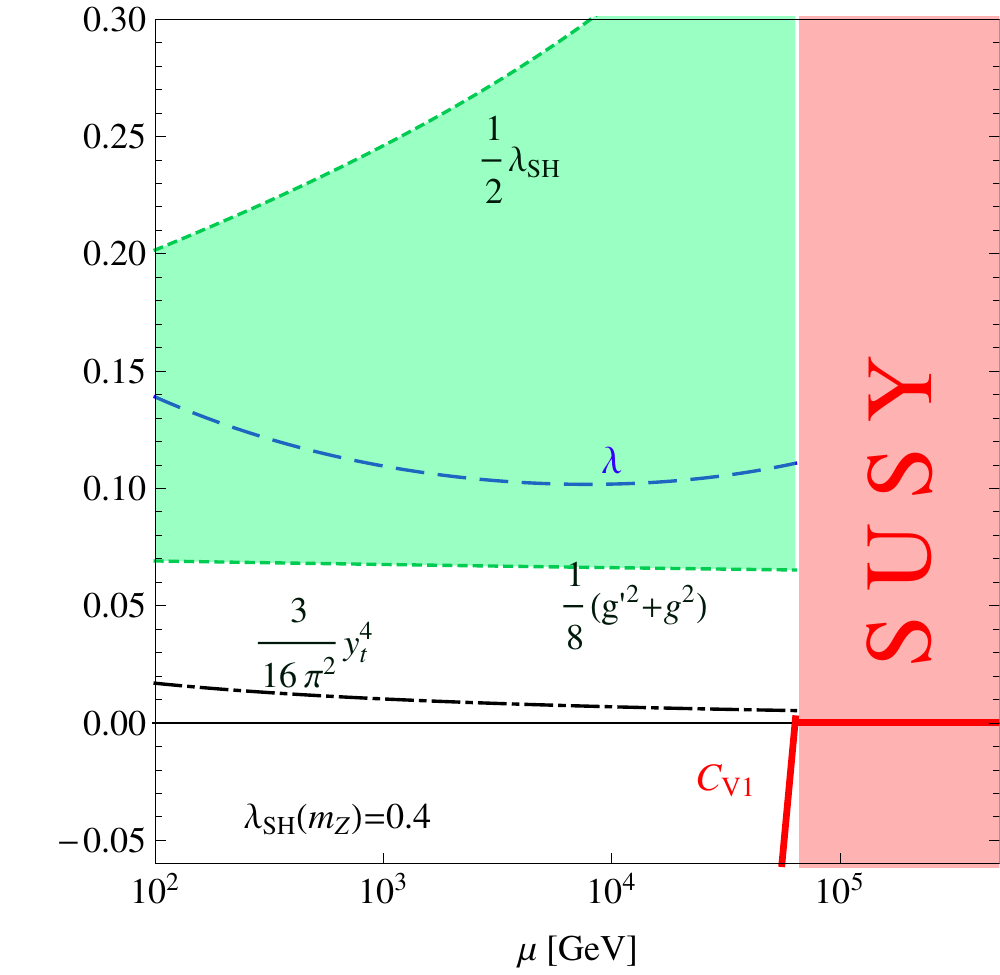}  \,\, \includegraphics[width=7.8cm]{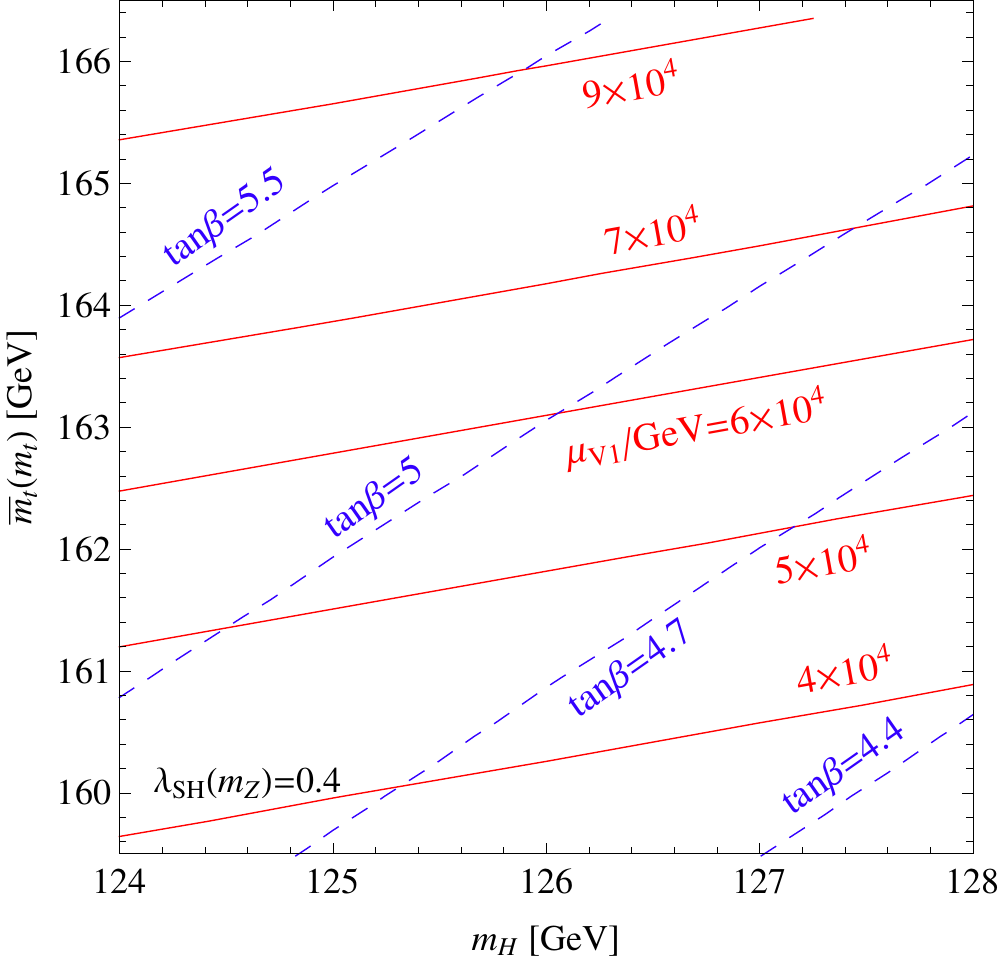} \vskip1cm
\includegraphics[width=7.8cm]{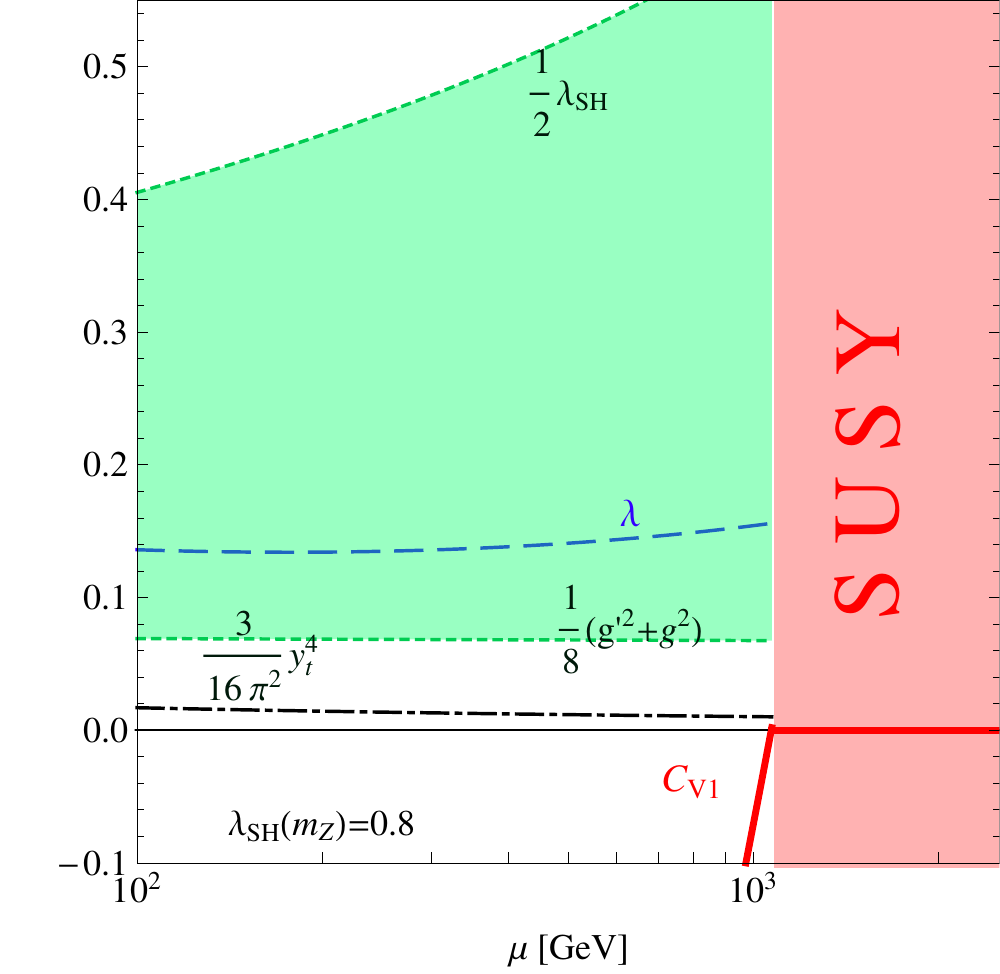} \,\,  \includegraphics[width=7.8cm]{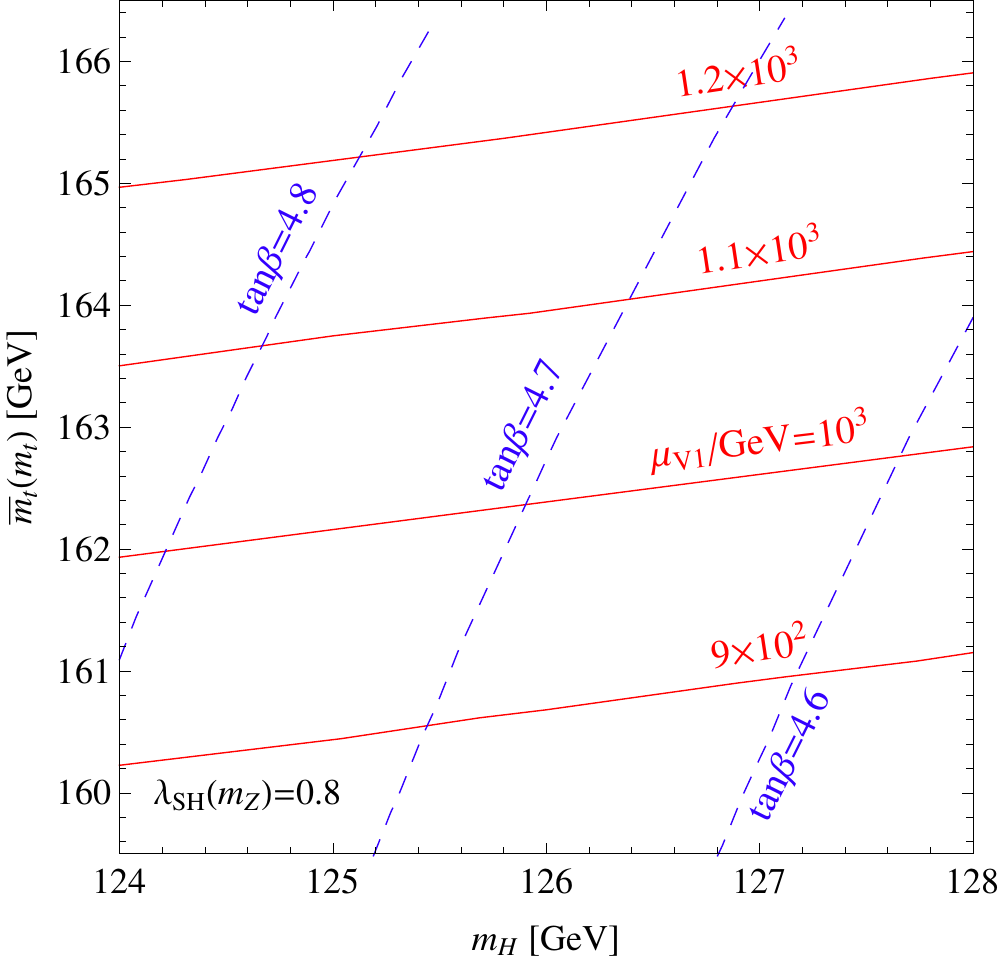}    
 \end{center}
\caption{\baselineskip=15 pt
Examples of viable merging with the NMSSM for two values of $\lambda_{SH}(m_Z)$: 0.4 (upper panels) and 0.8 (lower panels). In the left panels we show the running with the scale of the Veltman condition (\ref{VCsinglet}) as well as the running of $\lambda(\mu)$ and that of the couplings defining the interval (\ref{interval}): $\frac{1}{2}\lambda_{SH}(\mu)$ and $\frac{1}{8}[g^2(\mu)+
g«^2(\mu)]$. In the right panels we plot contour lines of constant $\mu_{V_1}$ and $\tan\beta$ 
in the plane $(\overline m_t(m_t),m_H)$  assuming $\alpha_3(m_Z) =0.1196$.}
\label{fig-VC1SMSsusy}
\vskip .1 cm
\end{figure}
For larger values of $\lambda_{SH}(m_Z)$ [$\lambda_{SH}(m_Z)\gtrsim 0.3$] the value of $\lambda(\mu_{V_1})$ enters the interval (\ref{interval}) and the merging becomes possible. This is shown in Fig.~\ref{fig-VC1SMSsusy}
where two different values of $\lambda_{SH}(m_Z)$ are used: 0.4 (upper panels) and 0.8 (lower panels). For each curve we again selected, as in fig.~\ref{fig-VC1SMSnonsusy}, the values of $\lambda_{S}(m_Z) $ and $\lambda_{f}(m_Z)$ such that the Veltman condition along the $S$ direction (\ref{VCS}) is satisfied, along with $\lambda_S(\mu_{V1})=0$. In particular for the upper panels we have chosen $\lambda_S(m_Z) \approx 0.03$ and $\lambda_f(m_Z) \approx1.1$, 
while in the lower panels $\lambda_S(m_Z) \approx0.05$ and  $\lambda_f(m_Z)\approx1.59$.
Plots in left panels show the RGE running of the different terms contributing to the VC1 where we can see that threshold corrections are rather tiny. In particular we can see that for $\lambda_{SH}(m_Z)=0.8$ the merging happens at scales $\sim$ TeV. Similarly the matching of $\lambda$ from Eq.~(\ref{matching}) provides (in the absence of mixing) the value of $\tan\beta$ at the scale $\mu_{V_1}$. In the right plots of Fig.~\ref{fig-VC1SMSsusy} we show, for the corresponding values of $\lambda_{SH}(m_Z)$, contour plots of the merging scale $\mu_{V_1}$ and $\tan\beta$ for $x_t=0$.Ä

\begin{figure}[h!]
\vskip .5cm 
 \begin{center}
\includegraphics[width=7.7cm]{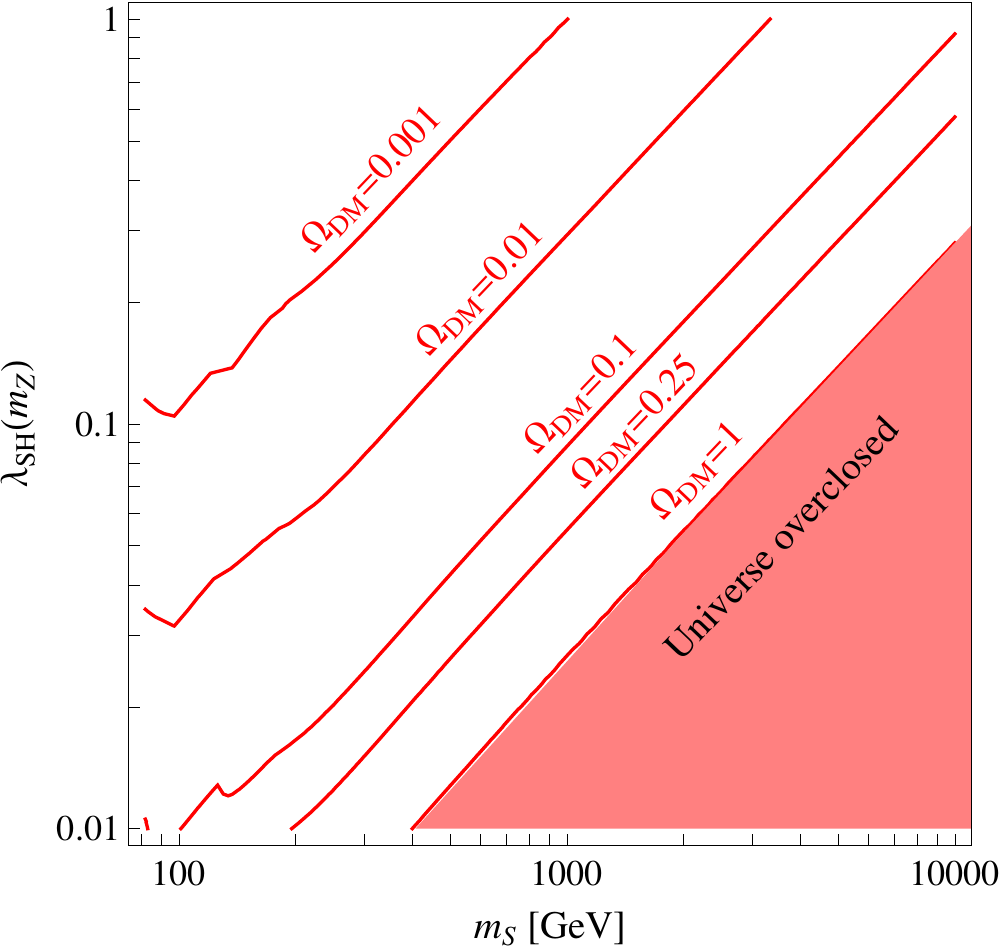} \,\,\,\,\, \includegraphics[width=7.6cm]{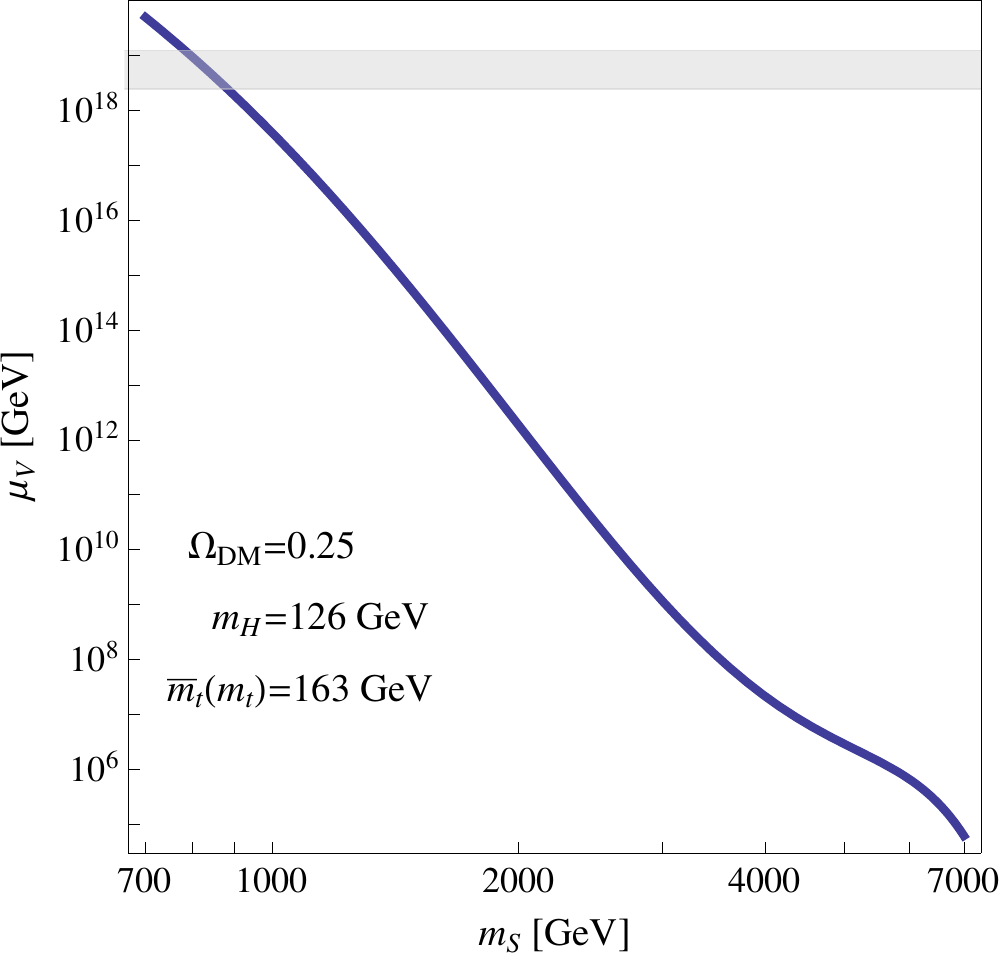} 
 \end{center}
\caption{\baselineskip=15 pt
Left panel: Contour levels of  $\Omega_{\rm DM}$ today. Note that, according to WMAP~\cite{PDG}, $\Omega_{\rm DM}\approx 0.25$.
Right panel: The scale where the Veltman condition is satisfied as a function of $m_S$ when $\Omega_{\rm DM}\approx 0.25$. We also 
fixed $m_H=126$ GeV and $\overline{m_t} (m_t) =163$ GeV.
}
\label{fig-DM}
\vskip .1 cm
\end{figure}

Finally notice that a spin-off of the model is that there are candidates to Dark Matter. In fact the scalar potential (\ref{potencialampliado}) has the discrete symmetry $\mathcal I\to -\mathcal I$ which opens up the possibility of the real scalar $\mathcal I$ as a Dark Matter candidate. In fact the Lagrangian term  $\lambda_2^2|S|^2|H|^2$ in (\ref{potencialampliado}) yields the contact interaction $\frac{1}{4}\lambda_2^2 \mathcal I^2 h^2$ and the tri-linear coupling $\frac{1}{2}\lambda_2^2v\mathcal I^2 h$ with the SM Higgs $h$ which provide annihilation amplitudes into the SM channels: $hh$, $WW$, $ZZ$, $tt$,\dots. This possibility was widely explored in the literature~\cite{DM} and the requirement of correct thermal cosmological abundance leads, for $\lambda_{2}=\mathcal O(1)$, to mass values $m_S=\mathcal O(1)$ TeV. More precisely we plot
in the left panel of Fig.~\ref{fig-DM} the contour levels  in the plane $(m_S,\lambda_{SH}(m_Z))$, including the contour corresponding to the thermal 
density $\Omega_{\rm DM}\approx 0.25$, as given by WMAP~\cite{PDG}.

If we cross the plot on the left panel of Fig.~\ref{fig-DM} with the information contained in Fig.~\ref{fig-muV1SMS} we can relate $\mu_V$ to $m_S$ as it is shown in the right panel of Fig.~\ref{fig-DM}. We can see that for $\mu_V\simeq M_{Pl}$ the correct thermal density is obtained for $m_S\simeq 1$ TeV and $\lambda_{SH}(m_Z)\simeq 0.05$, while for lower values of $\mu_V$ we obtain larger values of $m_S$. In particular for $\mu_V\sim 10^5$ GeV the correct thermal density is obtained for
$m_S\simeq 7$ TeV and $\lambda_{SH}(m_Z)\simeq 0.4$. Note that in the considered range of $m_S$ values the channels $h\to \mathcal I\mathcal I,\mathcal R\mathcal R$ is kinematically forbidden and there are no constraints from the invisible Higgs width. 
Of course one has to prevent the decay $\mathcal I\to ff$  from the fermion Lagrangian term  in (\ref{Lagrangianf}) $\mathcal I \psi^T i\gamma_5 \psi$ [where $\psi^T=(f,\bar f)$ is a four-dimensional Majorana fermion] which implies the condition $m_\mathcal I<2 m_f$. 

Notice that, after considering the constraints on the invisible Higgs decay, the region on the left of the $\Omega_{DM}=0.25$ curve in the left panel of Fig.~\ref{fig-DM} is allowed but the produced thermal density is too small and one would need another DM candidate. In particular the Majorana fermion $\psi$ itself is a candidate to Dark Matter through the coupling $\mathcal R \psi^T \psi$ as it annihilates into the Higgs field through the mixing of $\mathcal R$ with the physical Higgs $h$~\cite{Cheung:2012gi}. On the other hand the region on the right of the $\Omega_{DM}=0.25$ curve in the left panel of Fig.~\ref{fig-DM} is excluded as it would overclose the Universe.

\section{Conclusions}
\label{sec-concl}
The hierarchy problem for the Standard Model as an effective theory below a given cutoff $\Lambda$ is twofold:
\begin{itemize}
\item
On the one hand the presence of quadratic divergences makes the Higgs mass quadratically sensitive to the cutoff scale $\Lambda$. This is a purely Standard Model problem which is generated by the existence of quadratic divergences.
\item
On the other hand the Standard Model must be UV completed at the scale $\Lambda$ by a theory without quadratic divergences. Then:
\begin{itemize}
\item
Either there is no Higgs in the UV theory because the low energy Higgs is composite as a consequence of some infrared strong dynamics and dissolves at high energy scales.
\item
Or quadratic divergences identically cancel in the UV theory, as it is the case of a supersymmetric theory even in the presence of soft breaking terms.
\end{itemize}
Still the matching of the parameters of the high energy and the low energy theories should guarantee light Higgs mass parameters at the merging scale. This requires a fine-tuning on the UV theory parameters whose responsibility entirely lies in the high-energy theory. Of course this fine-tuning could be avoided if the cutoff $\Lambda$ is only a loop factor larger than the electroweak scale.
\end{itemize}

As it was already pointed out the absence of quadratic divergences -- dubbed Veltman condition -- is automatically satisfied in supersymmetric theories. However the absence of experimental hints of supersymmetric partners~\cite{SUSYexp} seems to imply that supersymmetry, if it exists at all, might be realized at high enough scales $\Lambda$ in which case requiring the absence of quadratic sensitivity on the cutoff of the low energy effective theory implies that the matching between the supersymmetric and non-supersymmetric theories should be done at the scale at which the Veltman condition is satisfied. In this case the low-energy effective theory should not exhibit any quadratic sensitivity on the cutoff scale. 

In this paper we have explored the general consequences of imposing the absence of quadratic divergences in the Standard Model and extensions thereof, a condition first imposed by Veltman in the context of the Standard Model. This should provide a solution to the above first step of the hierarchy problem. Does this imply a full solution to the hierarchy problem? The answer is clearly no, as the matching at the Veltman scale requires a fine-tuned relationship between parameters (step two above) to have a light Higgs squared mass parameter at the merging scale. However this opens up the possibility that this relation be explained within the high-energy theory: either we admit that we do not have any explanation for this fine-tuning, or it might have an environmental selection origin or perhaps it is due to some symmetries or properties of the UV theory. In this way the fulfillment of the Veltman condition allows us to postpone the solution of the hierarchy problem on the knowledge of the UV theory. In other words, our ignorance on the latter, or the absence of BSM experimental signatures, would prevent us from solving the whole hierarchy problem: a patently obvious truth.   

We have considered two models for the low energy effective theory: the Standard Model and its extension with a light complex scalar and a light fermion. The results within the Standard Model point towards a merging with the MSSM around the Planck scale and a value of $\tan\beta$ in the range $1<\tan\beta<2$. However for the central values of the top quark and Higgs masses and strong coupling $\alpha_3(m_Z)$ the merging happens at slightly trans-Planckian scales putting doubts on the consistency of the theory as gravitational effects are not considered. 

Furthermore we have shown that extending the Standard Model with light bosonic and fermionic degrees of freedom one could decrease the matching scale. In particular we have studied the Standard Model extended by a complex singlet, coupled to the Higgs field, with a coupling $\lambda_{SH}$, and to a singlet fermion, merging at the Veltman scale with its supersymmetric extension. In this case depending on the value of the coupling $\lambda_{SH}$ we can lower the matching scale towards the TeV scale and get a prediction for $\tan\beta$ in the supersymmetric merging theory. In particular for values of $\lambda_{SH}$ such that the Veltman scale is $\mathcal O$(few) TeV we can predict that in the supersymmetric theory $\tan\beta\simeq 4-5$.

Finally we have observed that a spin-off of the model is that it contains possible candidates to Dark Matter: in particular the real scalar $\mathcal I\equiv\textrm{Im}S$, which has the discrete symmetry $\mathcal I\to -\mathcal I$, and the Majorana fermion $\psi$ where $\psi^T=(f,\bar f)$, $f$ being the Weyl fermion in the supermultiplet $\mathcal T$. We postpone a more detailed analysis of the Dark Matter capabilities of the model to future work.


\vskip 1 cm

\section*{Acknowledgments}
The work of MQ is partly supported by the Spanish Consolider-Ingenio 2010 Programme CPAN (CSD2007-00042) and by CICYT-FEDER-FPA2011-25948. MQ would also like to acknowledge discussions with A.~Delgado on the meaning of quadratic divergences. We thank the Galileo Galilei Institute for Theoretical Physics for hospitality and the INFN for partial support during the completion of this work.




\end{document}